\documentclass[12pt]{article}

\usepackage{amsmath}            
\usepackage{amssymb}            %
\usepackage{amsfonts}           %
\usepackage{amsthm}             %

\usepackage{graphicx}           
\usepackage[utf8]{inputenc}     
\usepackage{aas_macros}				  
\usepackage{bm}                 
\usepackage[nopatch=footnote]{microtype}          
\usepackage{etoolbox}           
\usepackage[T1]{fontenc}        

\usepackage{slashed}				
\usepackage{mathrsfs}				
\usepackage{bbm}					  
\usepackage{cancel}					
\usepackage[normalem]{ulem} 
\usepackage{youngtab}	    	
\usepackage{mleftright}     

\usepackage[dvipsnames]{xcolor}
\usepackage[hang,flushmargin]{footmisc} 

\usepackage{fancyhdr}		
\usepackage{lipsum}			
\usepackage[most]{tcolorbox} 
\usepackage{subcaption}	
\usepackage{cite}			  
\usepackage{wrapfig}    

\usepackage{booktabs}		
\usepackage{nicefrac}		
\usepackage{multirow}		

\usepackage[font=small]{caption} 	
\usepackage{float}         			  
\usepackage{lineno}               
\usepackage{ccicons}              

\usepackage[margin=2.5cm]{geometry} 
\usepackage{changepage}             
\numberwithin{equation}{section}    
\usepackage{marginnote}             

\usepackage{multicol}
\usepackage{relsize}
\patchcmd{\thebibliography}
  {\list}
  {\begin{multicols}{2}\smaller\list}
  {}
  {}
\appto{\endthebibliography}{\end{multicols}}

\let\oldenumerate\enumerate
\renewcommand{\enumerate}{
  \oldenumerate
  \setlength{\itemsep}{4pt}
  \setlength{\parskip}{0pt}
  \setlength{\parsep}{0pt}
}

\let\olditemize\itemize
\renewcommand{\itemize}{
  \olditemize
  \setlength{\itemsep}{4pt}
  \setlength{\parskip}{0pt}
  \setlength{\parsep}{0pt}
}

\newcommand{\email}[1]{\href{mailto:#1}{#1}}

\fancypagestyle{firststyle}{
  \rhead{\footnotesize%
  \texttt{\FlipTR}%
  }}

\usepackage{etoc}

\usepackage{anyfontsize} 
\usepackage{scalefnt}       
\newcommand\acro[1]{{\scalefont{.93}{#1}}}

\renewcommand{\tilde}{\widetilde}         
\renewcommand{\text}{\textnormal}	        

\newcommand{\SO}[1]{\ifmmode
  \textnormal{\acro{SO(}}#1\textnormal{\acro{)}}
  \else \acro{SO($#1$)} \fi}

\newcommand{\SU}[1]{\ifmmode
  \textnormal{\acro{SU(}}#1\textnormal{\acro{)}}
  \else \acro{SU($#1$)} \fi}

\newcommand{\Sp}[1]{\ifmmode
  \textnormal{\acro{Sp(}}#1\textnormal{\acro{)}}
  \else \acro{Sp($#1$)} \fi}

\usepackage{undertilde} 

\newcommand{\inv}{^{-1}}

\newcommand{\D}[2][]{\ensuremath{\operatorname{d}\mkern-3mu^{#1}\mkern-1mu{#2}}\,}
 
\usepackage{scalerel} 

\usepackage{pifont}
\newcommand\Vtextvisiblespace[1][.3em]{%
  \mbox{\kern.06em\vrule height.3ex}%
  \vbox{\hrule width#1}%
  \hbox{\vrule height.3ex}}

\usepackage{xparse}
\ExplSyntaxOn
\NewDocumentEnvironment{nolabel}{}{
  \cs_set_eq:NN \label \use_none:n
  \cs_set_eq:cN { ltx@label} \use_none:n
}{}
\ExplSyntaxOff

\usepackage[light]{FiraSans}

\usepackage{makecell}
\usepackage{wrapfig}

\usepackage{framed} 

\newcommand{\Ha}{{\text{\acro{H}}}} 
\newcommand{\HH}{{\text{\acro{H}$_2$}}}

\newcommand{\Hm}{{\text{\acro{H}$^-$}}}

\newcommand{\Hal}{{\text{\acro{H$\alpha$}}}} 

\newcommand{\Jtwo}{\text{J}_{21}}

\newcommand{\DeltaC}{\Delta_\text{c}}
\newcommand{\tauHub}{\tau_\text{Hub}}

\newcommand{\popiii}{{\scalefont{.95}{Pop~III}}}

\usepackage[
	colorlinks=true,
	citecolor=green!50!black,
	linkcolor=NavyBlue!75!black,
	urlcolor=green!50!black,
	hypertexnames=false]{hyperref}

\usepackage{orcidlink}			
\usepackage{cleveref}
\crefformat{equation}{(#2#1#3)}	
\crefrangeformat{equation}{(#3#1#4\,--\,#5#2#6)} 

\graphicspath{{figures/}}       

\begin{document}

\newcommand{\FlipTR}{UCR-TR-2025-FLIP-BSG-75, CETUP2025-014} 
\thispagestyle{firststyle}


\begin{center}
    {\LARGE \textbf{Direct Collapse Black Hole Candidates\\from Decaying Dark Matter} \par}
    \vskip .5cm




\newcommand{\authorA}{Yash Aggarwal}
\newcommand{\emailA}{yagga003@ucr.edu}
\newcommand{\orcidA}{0000-0002-3862-0622}
\newcommand{\institutionA}{
		Department of Physics \& Astronomy, 
	    University of  California, Riverside, 
	    \normalfont{CA} 92521}

\newcommand{\authorB}{James B. Dent}
\newcommand{\emailB}{jbdent@shsu.edu}
\newcommand{\orcidB}{0000-0001-9417-1297}
\newcommand{\institutionB}{
		Department of Physics, Sam Houston State University, %
		Huntsville, \acro{TX} 77341}

\newcommand{\authorC}{Philip Tanedo}
\newcommand{\emailC}{flip.tanedo@ucr.edu}
\newcommand{\orcidC}{0000-0003-4642-2199}

\newcommand{\authorD}{Tao Xu}
\newcommand{\emailD}{tao.xu@ou.edu}
\newcommand{\orcidD}{0000-0002-1193-8470}
\newcommand{\institutionC}{
		Department of Physics \& Astronomy, University of %
		Oklahoma, Norman, \acro{OK} 73019}


\begin{center}
	\textbf{\authorA}$^{a}$,
	\textbf{\authorB}$^{b}$,
	\textbf{\authorC}$^{a}$, and
	\textbf{\authorD}$^{c}$
	\par

	\texttt{\footnotesize \email{\emailA}}~\orcidlink{\orcidA},
	\texttt{\footnotesize \email{\emailB}}~\orcidlink{\orcidB},
	\texttt{\footnotesize \email{\emailC}}~\orcidlink{\orcidC},
	\texttt{\footnotesize \email{\emailD}}~\orcidlink{\orcidD}
\end{center}

\begin{quotation}\noindent
	\footnotesize
	\noindent$^{a}$
	\textit{\institutionA} 
	\\\noindent $^{b}$ \textit{\institutionB} 
	\\\noindent $^{c}$ \textit{\institutionC} 
\end{quotation}

\end{center}

\begin{abstract}
\noindent
Injecting 1--13.6~eV photons into the early universe can suppress the molecular hydrogen abundance and alter the star formation history dramatically enough to produce direct collapse black holes. These, in turn, could explain the recently observed population of puzzling high-redshift supermassive black holes that appear to require super-Eddington accretion. 
We show that axion dark matter decay in the intergalactic medium can account for this energy injection. 
We use a single zone model of the gas core and semi-analytically evolve its chemo-thermal properties to track the conditions for which the system becomes an atomic cooling halo---a necessary precursor for the production of heavy black hole seeds to explain the high-redshift black hole population.
Windows of axion masses between 24.5--26.5\,eV with photon couplings as low as $4\times 10^{-12}\,\text{GeV}\inv$ may realize this atomic cooling halo condition. We highlight the significance of the band structure of molecular hydrogen on the effectiveness of this process and discuss estimates of the heavy seed population and prospects for testing this model. 

\end{abstract}

\small
\setcounter{tocdepth}{2}
\tableofcontents
\normalsize


\section{Introduction}

The first pre-stellar halos at redshift $z\gtrsim 15$ are ideal laboratories for dark matter. They are large, chemically simple gas volumes that sit in a dark matter overdensity that is free from background starlight or stellar magnetic fields. The gas is highly sensitive to the presence of a small amount of molecular hydrogen which permits it to cool, fragment, and collapse into the first (\popiii{}) stars. Dark matter can alter this gas chemistry and lead to dramatically observable changes in the star formation history. 
Injecting $1\text{--}13~\text{eV}$ photons into this environment can suppress the molecular hydrogen abundance. This allows the baryonic gas to heat to $T=10^4\,\text{K}$, at which point the system is able to cool through atomic hydrogen line emission. The gas may then collapse isothermally without fragmentation, and seed direct collapse black holes~\cite{Bromm:2001bi, Oh:2001ex, Inayoshi:2019fun, Smith:2019nuc, Prole:2023toz} that, over cosmic time, grow so massive that they would otherwise appear to require super-Eddington accretion to have formed in the standard pathway~\cite{Bromm:2002hb}.

In fact, astronomers have observed supermassive black holes in high-redshift quasars for more than a decade~\cite{Regan:2024wsu, Inayoshi:2019fun}. A surprisingly large population has recently been observed by \acro{JWST}, including \acro{UHZ1}~\cite{Natarajan:2023rxq} and \acro{GHZ9}~\cite{Kovacs:2024zfh}, $10^{7\text{--}8}\,{\rm M}_\odot$ black holes at $z=10$. 
Further hints may come from the observation of the so-called \emph{little red dots},  which may be compact starburst galaxies~\cite{Guia:2024toq, 2025ApJ...986..165T} or broad-line active galactic nuclei between $3.5 < z < 6.8$~\cite{2025ApJ...986..165T,2025ApJ...986..126K,Matthee:2023utn}.
The origin of these observed supermassive black holes is an open puzzle in cosmology~\cite{Banados:2017unc, Inayoshi:2019fun}. %

One possibility is that a source of Lyman--Werner band photons induces pre-stellar gas to monolithically collapse into a supermassive star. 
The supermassive star can then collapse to a black hole. An astrophysical source of such radiation could be halos that have already formed \popiii{} stars, which then irradiate their neighbors. 
However, the effects of \HH{} self-shielding as the gas reaches high densities implies that the critical flux may be unfeasibly large. 
Models where this Lyman--Werner flux originates from dark matter within the halo~\cite{Friedlander:2022ovf} can mitigate this effect with adiabatic contraction~\cite{Lu:2023xoi,Lu:2024zwa}, but decaying dark matter models must have a tuned mass relative to individual Lyman--Werner resonances.
On the other hand, recent simulations show that a weaker condition---the formation of atomic cooling halos at low gas densities---are likely to produce measurable differences in the black hole mass function and may be sufficient to produce heavy enough seeds for direct collapse~\cite{2019Natur.566...85W, Regan:2019vdf}.

In this paper, we show that $\mathcal O(20~\text{eV})$ mass decaying dark matter can produce atomic cooling halos and propose that this mechanism may explain the observed population of supermassive black holes. We go beyond recent treatments of this problem by accounting for the enhanced flux of decay products from the intergalactic medium (\acro{IGM}) as opposed to \emph{in situ} production within the halo, using the redshifted spectrum to populate a finite region of the Lyman--Werner band, and accounting for the effects of dynamical heating. We present the parameter space of axion dark matter to produce atomic cooling halos.

\subsection*{Prior Literature}

\paragraph{Review literature for supermassive black holes}
The 2019 review by Inayoshi et al.~\cite{Inayoshi:2019fun}, review article by Smith and Bromm \cite{Smith:2019nuc}, the 2018 Prato proceedings~\cite{Woods:2018lty}, and the 2025 review by Reagan and Volonteri~\cite{Regan:2024wsu} are accessible entry points to the astrophysics of supermassive black holes and their assembly.\footnote{Ref.~\cite{Natarajan2018-xt} is a particularly good public-level introduction that may even be accessible to particle physicists.}

\paragraph{Difficulty of Light Seeds}
Several methods have been suggested in the literature to explain the existence of high-$z$ \acro{SMBH}s.
The straightforward way is to grow $\sim 10\text{--}100 \, {M}_\odot$ \popiii{} \emph{light seeds} through accretion.
However, to get to observed masses of almost $\gtrsim 10^6 \, {M}_\odot$ by $z \gtrsim 6$, light seeds require sustained and near Eddington limited growth throughout the age of the universe \cite{Woods:2018lty, Smith:2019nuc, Inayoshi:2019fun, Regan:2024wsu}. This is challenging since
their host galaxies provide shallow potential wells \cite{Inayoshi:2019fun, Klessen:2023qmc}, and the strong radiative feedback drives the gas away from the surrounding environment \cite{Woods:2018lty, Smith:2019nuc, Inayoshi:2019fun, Regan:2024wsu}. 
A possible solution around these restrictions are episodes of super-Eddington growth that may achieve observed masses \cite{Smith:2019nuc, Inayoshi:2019fun, Regan:2024wsu}.

\paragraph{Heavy Seed Scenario} Alternatively, \emph{heavy seeds} with initial masses of $\sim 10^3 - 10^5 \, {M}_\odot$ have become attractive candidates to explain the high-redshift \acro{SMBH} abundance. 
The heavy seed pathway includes a class of mechanisms that are not entirely distinct from each other.
One plausible path is runaway mergers of stars in dense clusters that give rise to $\sim 10^3 \, {M}_\odot$ seeds \cite{Inayoshi:2015yqa,Smith:2019nuc, Inayoshi:2019fun, Klessen:2023qmc, Regan:2024wsu}. Another possibility, and focus of this paper, includes formation of $\sim 10^3\text{--}10^5 \, {M}_\odot$ \emph{supermassive stars} that collapse directly into a black hole, without a supernovae feedback. The \emph{Direct Collapse Black Hole} (\acro{DCBH}) scenario requires suppression of \HH{} cooling in pristine metal free halos through irridiation with UV fllux, high baryon streaming velocities, or heat generated through rapid halo mergers and accretion.

\paragraph{Direct Collapse}
Of the multiple proposed mechanisms to form supermassive black holes, we focus on producing massive seeds through direct collapse of a pre-stellar gas. In order to realize this, the gas must avoid early fragmentation into \popiii{} stars and become atomic cooling halos. Simulations indicate that these halos may then undergo direct collapse.
An early manifestation of this direct collapse pathway was proposed by Loeb and Rasio~\cite{Loeb:1994wv} in 1994 and by Bromm and Loeb in 2002~\cite{Bromm:2002hb}. 
A key challenge is to model the angular momentum transport to determine whether the gas forms sufficiently massive seeds. Lodato and Natarajan found that direct collapse could indeed form sufficiently heavy seeds~\cite{Lodato:2006hw,Lodato:2007bs} (see also Ref.~\cite{Koushiappas:2003zn, Begelman:2006db}). They showed that angular momentum initially supports a pre-galactic disk against fragmentation, and is then transported away by gravitational instabilities so that the gas collapses without fragmentation, thereby forming $\sim 10^5 \,M_\odot$ seeds. 

The bottleneck for direct collapse is the presence of molecular hydrogen, \HH{}, in the gas. \HH{} is the primary coolant for cool, low-metallicity gas and is responsible for the fragmentation that produces \popiii{} stars rather than atomic cooling halos. 
The standard astrophysics proposal to remove \HH{} is to assume an external flux of Lyman--Werner radiation from nearby stars. These $\mathcal O(10\,\text{eV})$ photons dissociates molecular hydrogen. The critical flux of Lyman--Werner radiation, $J_\text{crit}$, can vary by orders of magnitude depending on the simulation, though it is typically large compared to what one expects from the distribution of the first stars.

In addition to Lyman--Werner radiation, lower-energy $\mathcal O(2-10\,\text{eV})$ photons can suppress \HH{} abundance by destroying the \Hm{} ions that are catalysts for \HH{} formation at low temperatures. Thus the minimum flux to induce direct collapse is usually presented as a \emph{critical curve} on the plane of photodissociating and photodetatching flux~\cite{Wolcott-Green:2011tul, Wolcott-Green:2016grm}. Beyond this critical flux, the gas is said to enter a \emph{zone of no return} after which its evolution is expected to reach a  $10^5\,M_\odot$ supermassive star by $z\sim 15$~\cite{Omukai:2008wv} that directly collapses into a black hole, as understood over a series of papers in Refs.~\cite{Bromm:2002hb, Oh:2001ex, Omukai:2000ic, Begelman:2006db, Lodato:2006hw, Lodato:2007bs} and reviewed in ~\cite[Sec.~5]{Inayoshi:2019fun}. We review the status of simulations in Section~\ref{sec:simulations:for:collapse}.

\paragraph{Dark Matter Energy Injection}
Friedlander, Schön, and Vincent recently examined the possibility that dark matter could provide the Lyman--Werner radiation for direct collapse~\cite{Friedlander:2022ovf}. They found a mild tension between the dark matter decay or annihilation yield compared to the critical flux required for direct collapse. 
A key challenge is that \HH{} self-shields against external fluxes.\footnote{In a dense gas, \HH{} can efficiently absorb Lyman--Werner photons and reform. This allows \HH{} to build up within the volume, which in turn further contributes to shielding} 
On the other hand, Kusenko, Lu, and Picker pointed out that this effect is mitigated by the adiabatic contraction~\cite{Blumenthal:1985qy} of the dark matter density in response to the collapse of baryonic matter~\cite{Lu:2024zwa} (see also their embedding into a Majoron model, Ref.~\cite{Lu:2025kbe}). 
This highlights a difference between the \emph{in situ}\footnote{As Aaron Vincent describes it, ``The call is coming from inside the house.''} production of Lyman--Werner photons versus the case of external irradiation from nearby stars: the scaling of the \emph{in situ} flux with the dark matter density can naturally mitigate self-shielding. 

One limitation of these scenarios is that the in situ dark matter decays produce narrow lines in the photon spectrum. The Lyman--Werner band is a series of thin lines that correspond to excitations between rotational--vibrational (rotovibrational) states of \HH{}. Even if we assume the dark matter mass is tuned to be near one of these lines, with effectively a single frequency, the dark matter decay spectrum would only be able to dissociate the fraction of the \HH{} population that exists in a given rotovibrational state. 

In a study unrelated to direct collapse, Qin~et\ al.\ show that the energy injected from dark matter decays in the intergalactic medium can be much larger than that from within the halo~\cite{Qin:2023kkk}. This is due to the volume enhancement for a shell of dark matter that is a far away. Emission from this shell is redshifted and broadens the energy spectrum so that it may span a finite slice of the Lyman--Werner band. This further addresses self-shielding because it irradiates the volume \emph{before} the halo collapses and can suppress any initial \HH{} formation. 

In this paper, we present the viability for this \emph{intergalactic} population of decaying dark matter to induce the conditions for direct collapse black holes. 
%
Other recent mechanisms to induce direct collapse black holes from new physics include Hawking radiation from a sub-population of primordial black hole dark matter~\cite{Lu:2023xoi}, direct seeding from primordial black holes~\cite{Kawasaki:2012kn}, superconducting cosmic string decay~\cite{Cyr:2022urs}, parametric resonances from ultralight dark matter~\cite{Jiao:2025kpn}, or enhanced density fluctuations at very high redshifts~\cite{Qin:2025ymc}. 
Earlier new physics mechanisms for supermassive black hole formation include self-interacting dark matter~\cite{Pollack:2014rja} (see also recent manifestations \cite{Feng:2020kxv, Xiao:2021ftk}) and the dark star scenario where $\mathcal O(100~\text{GeV})$ dark matter annihilation provides thermal support for baryonic clouds~\cite{Spolyar:2007qv,Natarajan:2008db,Spolyar:2009nt} and can heat the gas to overwhelm \HH{} cooling~\cite{Banik:2016qww}. 
The stellar dynamics of dark matter energy injection at cosmic dawn was studied in earlier work by Mack, Schön, Wyithe and collaborators in Refs.~\cite{Schon:2014xoa,Schon:2017bvu}; Qin and the \texttt{DarkHistory} collaboration~\cite{Liu:2023fgu,Qin:2023kkk,Liu:2023nct,Qin:2024fvr}; and Hou and Mack~\cite{Hou:2024hcj}.

\subsection*{This Paper is Organized as Follows}

Given the interdisciplinary nature of this topic, we provide background on astrophysics, \HH{} chemistry, and our dark matter model in Section~\ref{sec:background} for particle physicists with limited astrophysics background. This section may be skipped by experienced readers.
Section~\ref{sec:methodology} articulates our methodology and Section~\ref{sec:results} presents our results. Our viable axion parameter space is summarized in Fig.~\ref{fig:Axion.DCBH}. We conclude in Section~\ref{sec:conclusion} with a summary and discussion of further directions. 
To be as self-contained as possible, we provide appendices on our halo model, a comparison of the intergalactic medium to local photon contributions, and a review of molecular hydrogen and photodissociation.

\section{Astrophysics Background}
\label{sec:background}

We investigate the possibility that dark matter decay products  suppress molecular hydrogen and thus induce direct collapse black holes in the early universe. In this section, we present the astrophysical background to address the key question: \emph{what are the necessary conditions to produce direct collapse black holes}?  In the last subsection we review the our benchmark  axion dark matter model.
For a more extensive background on first stars, we refer the reader to the textbook by Loeb \& Furlanetto~\cite{Loeb&Furlanetto:2013} and the recent review by Klessen and Glover~\cite{Klessen:2023qmc}. 
For a background on supermassive black holes, we refer to the reviews in Refs.~\cite{Inayoshi:2019fun,Woods:2018lty,Smith:2019nuc,Regan:2024wsu}.

\subsection{Glossary of Terms}

\begin{itemize}
    \itemsep1em 
    \item[] \textbf{\popiii{} seeds} or \textbf{light seeds}: Light black hole seeds, $\sim 10 - 100 \, M_\odot$, formed out of first generation of \popiii{} stars in $\sim 10^5 - 10^6 \ M_\odot$ minihalos as a result of \HH{} cooling. These black holes are prone to stunted growth due to strong radiative feedback from their hot progenitor stars (surface temperatures near $10^5 \, \text{K}$), supernova feedback, and presence in shallow gravitational potential wells.

    \item[] \textbf{Direct collapse black holes} (\acro{DCBH}s) or \textbf{heavy seeds}: Initial black hole seeds of $\gtrsim 10^3 \ M_\odot$ born out of short-lived supermassive stars.\footnote{%
        To the best of our knowledge, there is no uniform definition for \emph{heavy seed}; we adopt the standard by recent papers by Regan and collaborators~\cite{Regan:2024wsu, Prole:2024koa, OBrennan:2025sft}, though we note that earlier papers define a heavy seed to be between $10^4$--$10^6$\,$M_\odot$, e.g.~\cite{Pacucci:2019baa, 2021NatRP...3..732V}.
    }  
    Compared to \popiii{} stars, the supermassive stars are embedded in deeper gravitational potential wells, experience little radiative feedback due to cold surface temperatures ($\sim 10^4 \, \text{K}$ see Ref.~\cite{2018MNRAS.474.2757H}), and collapse into a black hole without undergoing a supernova\footnote{There are some exceptions: stellar masses near $55,000 \, M_\odot$ that undergo helium burning, rotating stars heavier than $10^6 \, M_\odot$, and non-accreting metal enriched $\sim 500,000 \, M_\odot$ stars~\cite[Sec.~5.4.2]{Woods:2018lty}.}.
    
    \item[] \textbf{Atomic cooling halos}: Halos with no history of \HH{} cooling, star formation, or metal enrichment that reach a gas temperature of $T \approx 10^4 \, \text{K}$. Hydrogen line emissions allow the gas to cool in a way that realizes the necessary conditions for heavy seeds.

    \item[] \textbf{Monolithic collapse}: Atomic cooling halos whose gas cloud undergoes an isothermal collapse without violent gas fragmentation. This occurs when cooling is dominated by atomic hydrogen and \HH{} is suppressed through most of the star formation phase. This is understood to produce heavy seeds of order $\sim 10^5 \, M_\odot$.\footnote{Vigorous fragmentation near the end of star formation may reduce the seed masses~\cite{Prole:2023toz}.}

    \item[] \textbf{Direct collapse black hole candidates} (\acro{DCBH} candidates): 
    Halos that reaches the atomic cooling threshold, without necessarily requiring monolithic collapse. 
    Recent simulations show that this may be a sufficient condition to form heavy seeds at least as massive as $\sim 10^3$--$10^4 \, M_\odot$~\cite{2019Natur.566...85W,Regan:2019vdf,Regan:2020drm, Prole:2024koa, Mone:2024yxl}. We choose this to be our benchmark for achieving novel astrophysical phenomenology from dark matter decay; see Section~\ref{sec:simulations:for:collapse}.
\end{itemize}

\subsection{Atomic Cooling and DCBH Candidates}
\label{sec:conditions.dcbh}

We investigate the formation of \acro{DCBH} candidates out the of pristine, metal-free galaxies that would otherwise form the first generation of stars (\popiii{} stars).
Simulations observe two necessary conditions to create heavy seeds instead of \popiii{} stars (see, e.g.~\cite{Inayoshi:2019fun}): 
\begin{enumerate}
    \item\label{heav:seed:condition:1} 
    Rapid gas accumulation 
    ($\gtrsim 0.1\,M_\odot/\text{yr}$) 

    \item\label{heav:seed:condition:2} 
    Suppression of \HH{} to prevent gas fragmentation. 
\end{enumerate} 
A necessary condition to meet these criteria is for the system to become an \emph{atomic cooling halo}.
In the absence of an appreciable \HH{} population, the gas in a would-be \popiii{}-star-forming halo tracks its virial temperature~\cite{Bromm:2001bi}. The gas grows denser and hotter until it reaches a temperature of $T \approx 10^4\,\text{K}$ when the virial mass reaches $M_\text{vir} \approx 10^7\,{M}_\odot$. 
At this temperature, thermal collisions excite the Lyman levels of \Ha{} atoms and permit Lyman emission to cool the gas. The gas then collapses quasi-isothermally\footnote{The Lyman emission rate scales with the temperature of the gas. If the gas is hot, more hydrogen atoms have the collisional energy to excite \Ha{} lines. If the gas is cold, there is less cooling and dynamical heating warms the gas toward its virial temperature.} and rapidly enough to satisfy the first heavy seed condition~\cite{Woods:2018lty}.
The accretion rate scales as~\cite[eq.~7]{Begelman:2006db}, %
\begin{align}
    \dot{M}_\text{acc} 
    \sim 
    \frac{ M_\text{J} }{ t_\text{ff} } 
    \sim 
    \frac{ c_\text{s}^3 }{ G_\text{N} } 
    \propto
    T^{3/2} \ ,
\end{align}
where $M_\text{J}$ is the Jeans mass, and $t_\text{ff}$ is the free-fall time, $c_\text{s}$ is the sound speed and $G_\text{N}$ is the gravitational constant. We refer to Refs.~\cite[eq.~5]{Klessen:2023qmc} and \cite[Tab.~3]{Inayoshi:2019fun} for reviews.  
The $T^{3/2}$ scaling connects the high temperatures realized by an atomic cooling halo to its rapid accretion rate. 
Reaching the atomic cooling limit suppresses fragmentation in first star forming gas clouds~\cite{2018MNRAS.475.4636R}, addressing the second condition for heavy seeds. 
In this study we thus take the realization of an atomic cooling halo as our the target for the formation of \acro{DCBH} candidates.

\subsection{How much molecular hydrogen suppression?}
\label{sec:simulations:for:collapse}

\paragraph{Conservative scenario: monolithic collapse}
An atomic cooling halo that satisfies both of the heavy seed criteria through most of its history results in the \emph{monolithic collapse} of gas clouds, with seed masses nearly $\mathcal{O}(10^5 \, {M}_\odot)$~\cite{2012MNRAS.422.2539I,Fernandez:2014wia,Inayoshi:2014rda}. Such seeds attain the mass needed to explain high-redshift quasars without any additional astrophysical mechanisms, though the large gas densities in the core promote \HH{} formation and thus require a stronger photon flux needed to suppress the \HH{} population.
A standard benchmark for the critical flux is to reach the \emph{zone of no return} where an atomic cooling halo reaches a density of $n_\text{p}\sim 10^4/\text{cm}^3$ while maintaining a temperature $T \sim 10^4\,\text{K}$~\cite[Fig.~1]{2012MNRAS.422.2539I}; 
The critical Lyman--Werner flux to reach this zone of no return is $J\sim 10^3\,\Jtwo$, where $\Jtwo$ is a standard unit of mean intensity presented below in \eqref{eq:J21}. 
This benchmark is the standard used in the late 2010s to determine the critical Lyman--Werner and photodetachment dissociation rates for heavy seed formation, e.g.\ Refs~\cite{Sugimura:2014sqa, Becerra:2014xea, Wolcott-Green:2016grm, 2016MNRAS.459.4209A, 2020MNRAS.492.4917L}.

 \paragraph{Benchmark scenario: atomic cooling halos are sufficient} 
Recent simulations suggest that it may be sufficient to require that a halo reaches the atomic cooling condition when including the effects of the astrophysical environment, see e.g.\ Ref.~\cite{Mone:2024yxl}. This condition may be attained at modest gas densities, $n_\text{p} \sim \mathcal O(\text{cm}^{-3})$, and thus requires a much smaller photon flux. Unlike the conservative scenario, the weaker condition permits the \HH{} fraction to grow after reaching atomic cooling, potentially leading to subsequent gas fragmentation.
In the absence of additional dynamics, the halo can still produce seed masses of nearly $\mathcal{O}(10^{3\text{--}4}\,M_\odot)$ while requiring a significantly smaller Lyman--Werner flux.

However, the astrophysical environment can offset the Lyman--Werner flux required to prevent \HH{} cooling. 
For example, mergers can dynamically heat the gas~\cite{Yoshida:2003rw} and some simulations show that even non-isothermal conditions can maintain high accretion rates to encourage direct collapse~\cite{Latif:2015aaa}.
Refs.~\cite{2019Natur.566...85W, Regan:2019vdf}
found that even $\mathcal O(\Jtwo{})$, three orders of magnitude less than the `zone of no return', can produce supermassive seeds in atomic cooling halos when combined with a period of rapid growth from a merger.
A follow up paper argues that similar flux produces seed \emph{intermediate mass} black holes, up to $10^4\,{M}_\odot$, when accounting for turbulence and fragmentation~\cite{Regan:2020drm}.
The simulation in Ref.~\cite{Prole:2024koa} shows how mergers of atomic cooling halos may boost accretion rates; the recently discovered $\infty$-galaxy may be observational evidence for these dynamics~\cite{vanDokkum:2025idp}.
For this reason, in this study we focus on the evolution of halo up that reach the atomic cooling limit with no additional requirements for its subsequent evolution. This is a credible checkpoint for the impact of dark matter energy injection in proto-galaxies and is straightforward to assess semi-analytically. 
Simulations suggest that this population of \acro{DCBH} candidates produces intermediate-mass seeds that shape the distribution of high-redshift black holes, while a fraction may evolve into the extremely massive black holes recently observed.

\subsection{The Birth and Death of Molecular Hydrogen}
\label{eq:birth:and:death:of:H2}

The key for a halo to become a \acro{DCBH} candidate is that the molecular hydrogen fraction is suppressed during its early evolution.
The first star-forming gas clouds are primarily composed of atomic hydrogen, \Ha{}. A small fraction of molecular hydrogen, \HH{}, allows the gas to cool through inelastic collisions that excite the \HH{} electronic ground state and subsequent photon emission; see Ref.~\cite{Galli:2012rf} for a review.
The critical amount of \HH{} to begin cooling is modest,\footnote{%
$x_i$ is the ratio of the species $i$ number density to that of the total number of hydrogen nuclei (including \Ha{}, \Hm{}, and \HH{}). In astronomy, this is often written as $f_i$ for species and $x_\text{e}$ for the ionization fractions. We follow the notation of Ref.~\cite{Qin:2025ymc} and use $x_{\HH{}}$ for the molecular hydrogen abundance as a ratio to hydrogen nuclei. This helps disambiguate from the self-shielding fraction $f_\text{sh}$.
\label{foot:xi:ratio}
}
$x_\HH{} = n_{\HH{}}/n_\text{p} \sim 10^{-4}$
~\cite{Tegmark:1996yt,Bromm:2001bi, Yoshida:2003rw}. 

In the standard scenario, halos form enough \HH{} to efficiently cool the gas by the time the gas reaches a temperature of $T \sim 10^3\,\text{K}$. At that point, \HH{} cooling kicks in and rapidly cools the gas to $T \sim 200 \, \text{K}$. 
The gas then fragments into the first generation of $\mathcal{O}(10 \,\text{--}100\,{M}_\odot)$ \popiii{} stars~\cite{2011Sci...331.1040C}. We refer to Ref.~\cite{Klessen:2023qmc} for a compilation of \popiii{} mass ranges.

To induce the direct collapse mechanism, one must remove \HH{} until it is no longer a viable coolant in the gas; for a review of the chemistry of \popiii{}-star forming gases, see~\cite{Abel:1996kh,Galli:1998dh, Galli:2012rf}. 
The formation of $\HH{}$ in pre-stellar clouds is catalyzed by free electrons,
\begin{align}
    \Ha{} + \text{e}^- & \;\to\; \Hm{} + \gamma
    &
    \Hm + \Ha{} \;\to\; \HH{} + \text{e}^-  \ .
    \label{eq:h2:hm:form} 
\end{align}
This points to two mechanisms by which a source of photons can remove \HH{} from astronomical environments. 
First, one can prevent \HH{} production by \emph{photodetaching} the electron from the precursor \Hm{} ion,
\begin{align}
    \Ha{}^- + \gamma &\;\to\; \Ha{} +  \text{e}^- 
    &
    E_\gamma \in \left[\, 0.75\,\text{eV},\; 13.6\,\text{eV}\, \right]
    \ . 
    \label{eq:hm.dest}
\end{align}
Second, photons in the Lyman--Werner ($11.2$--$13.6~\text{eV}$) frequency band can \emph{photodissociate} molecular hydrogen by exciting an intermediate electronic excited state:
\begin{align}
    \HH{} + \gamma 
    & \;\to\; \HH{}^* \;\to\; 2\Ha{} 
    &
    E_\gamma \in \left[\, 11.2\,\text{eV},\; 13.6\,\text{eV}\, \right]\quad \text{(Lyman--Werner)}
    \ . 
    \label{eq:h2.dest}
\end{align}
The \HH{}$^*$ dissociative decay occurs $\sim 15 \%$ of the time~\cite[p.\,176]{Loeb&Furlanetto:2013}.
The Lyman--Werner band is actually a series of narrow, but numerous, lines between specific ro-vibrational excitations of the \HH{} ground state and those of the \HH{}$^*$ excited state. For an appreciable dissociation rate, one typically assumes a photon spectrum that covers at least a finite continuous region of the band that extends across many individual transitions.
In term symbol notation, the excited states are $\HH^* \in \text{B}^1\Sigma_u^+$ and $\text{C}^1 \Pi_u$; we review \HH{} chemistry in Appendix~\ref{AP:H2:molecule}.

Because photodissociation directly removes \HH{} molecules rather than their precursors, it is more effective than photodetachment for similar rate coefficients. While we include the effect of photodetachment in this work, the primary means of suppressing the \HH{} abundance is through photodissociation via Lyman--Werner radiation.

\subsection{Decaying Dark Matter and Axions}

\HH{}-suppressing radiation may originate from dark matter decay.\footnote{Other decay channels can produce photons in the final state. Ref.~\cite{Qin:2023kkk} examines the effect of dark matter energy injection on \HH{} evolution and the first stars with dark matter masses from 10\,keV to above 10\,TeV.} 
We assume:
\begin{enumerate}
    \item A single spin-0 particle, $a$, comprises the entire dark matter density.

    \item The dark matter particle decays into photon pairs, $a\to\gamma\gamma$, with interaction strength $g_{a\gamma\gamma}$. This has dimension of inverse energy in an interaction term in the Lagrangian of the form    $g_{a\gamma\gamma}aF_{\mu\nu}\tilde{F}^{\mu\nu}/4$.

    \item The mass of the dark matter $m_a$ is in the range 
            \begin{align}
              0.75\,\text{eV} 
              \;\leq\; 
              \frac{m_a}{2} 
              \;\leq\; 
              13.6\,\text{eV}
              \label{eq:flux.1}
            \end{align}
          so that the decay products either photodetach or photodissociate \HH{}.  Photons with less energy than \eqref{eq:flux.1} do not affect the \HH{} abundance, and those with more energy ionize hydrogen and are absorbed before reaching the target halo.

    \item The decay rate depends on the coupling and mass as $\Gamma_a\sim g_{a\gamma\gamma}^2 m_a^3$. This scaling follows from dimensional analysis.
\end{enumerate}
In a viable model, $\Gamma_a$ is large enough to induce atomic cooling halos, while also being slow enough that the dark matter is cosmologically long lived to comprise the total present dark matter density. 
The decay into two photons has a monochromatic spectrum,
\begin{align}
    \frac{ \D{N_\gamma} }{ \D{E} } 
    &= 
    2\, \delta\!\left( \! E - \frac{m_a}{2} \! \right) \ ,
    \label{eq:flux.2}
\end{align}
which poses a challenge: in order to effectively dissociate \HH{}, one must tune the axion mass to align with one of the 76 narrow resonances that excite the \HH{} ground state. The thermal motion of dark matter within a halo does not appreciably broaden the spectrum. We show below that redshifting of photons from the intergalatic medium sufficiently broadens the spectrum at the halo to cover multiple lines.

\paragraph{Axions and \acro{ALP}s} A template for this class of models is axion dark matter.  
Axions are the pseudo-Nambu--Goldstone bosons of a spontaneously broken, global \acro{U(1)} symmetry~\cite{Weinberg:1977ma,Wilczek:1977pj} called Peccei-Quinn symmetry~\cite{Peccei:1977hh,Peccei:1977ur}. They are a possible solution to the strong \acro{CP} problem---axion models that address this issue are often referred to as \acro{QCD} axions and include the well-known \acro{KSVZ}~\cite{Kim:1979if,Shifman:1979if} and \acro{DFSZ}~\cite{Zhitnitsky:1980tq,Dine:1981rt} models and their variants; see, for example, the review~\cite{DiLuzio:2020wdo}.
There also exists a larger class of axion-like particles, or \acro{ALP}s, that are pseudo-Nambu--Goldstone bosons and have similar interactions as \acro{QCD} axions, but are not associated with a solution of the strong \acro{CP} problem~\cite{Graham:2015ouw,Bauer:2017ris,Irastorza:2018dyq}. 

The decay rate for axions and related pseudoscalar particles is
\begin{align}
    \Gamma_a 
    &= 
    \frac{ 
        g_{a\gamma\gamma}^2 m_a^3 
        }{ 
        64 \pi 
        } 
\ .
\label{eq:decay_rate}
\end{align} 
The dark matter density scales with redshift as:
\begin{align}
    \rho_{a}(z) 
    & = 1.26 \times 10^3  (1+z)^3 
        \, \text{eV}\, \text{cm}^{-3}
        \ .
    \label{eq:axion:density}
\end{align}
Constraints on \acro{ALP}s in this range come from 
observations of the cosmic optical background~\cite{Carenza:2023qxh, Porras-Bedmar:2024uql, Nakayama:2022jza},
$\gamma$-ray attenuation~\cite{Bernal:2022xyi},
perturbations to the cosmic recombination history~\cite{Xu:2024vdn},
\acro{CMB} distortions~\cite{Bolliet:2020ofj}, 
and the heating of dwarf galaxies~\cite{Wadekar:2021qae}.

\paragraph{Generality}
For simplicity we refer to our dark matter particle as an axion, $a$, rather than an \acro{ALP}, though we assume no direct connection to the strong \acro{CP} problem.  
There are explicit examples of axion models that produce cosmological dark matter with the required properties~\cite{Eroncel:2022vjg,Chatrchyan:2023cmz} and there are extensions of the \acro{KSVZ} model that solve the strong \acro{CP} problem in this parameter space~\cite{DiLuzio:2020wdo}. 
In fact, our analysis applies generally to any decaying dark matter model that produces monochromatic photons. Models in which parity-even scalars decay into two photons must be finely tuned against quantum corrections, but only differ in the numerical prefactor of the decay rate~\eqref{eq:decay_rate}. The template model for fermonic decaying dark matter are sterile neutrinos which also produce monochromatic photons from $\nu' \to \gamma \nu$, though the typical mass for viable models is in the keV range and are thus too energetic for \HH{} dissociation~\cite{Kusenko:2009up,Drewes:2016upu,Abazajian:2017tcc,Boyarsky:2018tvu,Vogel:2025aut,Akita:2025txo}.

\section{Methodology: Dark matter versus molecular hydrogen}
\label{sec:methodology}

We perform a semi-analytic analysis to identify the conditions under which axion decay suppresses \HH{} cooling sufficiently for a benchmark halo to reach the atomic cooling limit. 
We track the chemical and thermal evolution of the baryonic gas at the core of the halo. 
For clarity in relation to astrophysical formulae, we explicitly write Planck's constant $h$ and the Boltzmann constant $k_\text{B}$, but otherwise use natural units.

\subsection{Criteria for reaching atomic cooling limit}
\label{sec:dcbh.condition}

To satisfy the \acro{DCBH} candidate condition in Section~\ref{sec:conditions.dcbh}, 
a halo must suppress its molecular hydrogen fraction below a critical value, $x_{\HH{},\text{crit}}$.
This must be maintained until the halo heats to a temperature $T_\Hal{}$ where it may cool through atomic line emission:
\begin{align}
    x_\HH{} &< x_{\HH{},\text{crit}}
    &
    \text{for}
    &
    &
    T &\leq T_\Hal{} \equiv 10^4\,\text{K} \ .
    \label{eq:win:condition}
\end{align}
The critical \HH{} fraction is determined by the requirement that the timescale for cooling is longer than the halo's characteristic timescale. 
A common choice for this timescale is the
free fall time, which is typically used to determine whether a gas is pressure supported or if it collapses~\cite{Visbal:2014nogo ,Loeb&Furlanetto:2013, Klessen:2023qmc}. It is the time for a static gas cloud to collapse under gravity in the absence of stabilizing forces, see e.g.~\cite[eq.~8]{Klessen:2023qmc},
\begin{align}
    \tau_\text{ff}(z) & = 
    \sqrt{
        \frac{ 3 \pi }{ 32 G \bar{\rho}(z) }
    } 
    &
    \bar{\rho} (z) 
    &= 
    \DeltaC{}\,
    \rho_{\text{m}0}
    (1+z)^3
     \ ,
    \label{eq:tauff_rhobar}
\end{align} 
where $\bar\rho(z)$ is the average halo density at redshift $z$, $\DeltaC{} \approx 200$ is the mean overdensity\footnote{The mean overdensity is the ratio of the virial mass to the mass of a ball of radius equal to the virial radius  and density equal to the critical density of the universe: $\DeltaC{} = M_\text{vir}\left[\tfrac{4}{3} \pi R_\text{vir}^3 \rho_\text{crit}\right]\inv$.} \eqref{eq:concentration:parameter}, and $\rho_{\text{m}0} = 1.5 \times 10^3 \ \text{eV} \ \text{cm}^{-3}$ is the present matter density. In Section~\ref{sec:timescales} we motivate a more conservative choice for the characteristic timescale, the Hubble time.

\subsection{Background Halo Model}
\label{sec:bg:halo}

The chemical dynamics of the baryonic gas occurs in the background of a growing halo. 
We use the \emph{mass accretion history} prescription to model the growth of a benchmark halo. This is a fit for the median evolution of the halo mass $M(z)$ as a function of its present mass~\cite{Correa:2014xma},
\begin{align}
    M(z) &= M_0 (1+z)^{p f({M_0})}\text{e}^{ - f(M_0) z} 
    &
    M_0 &= 5\times 10^9\,M_\odot 
    \ ,
    \label{eq:MAH}
\end{align}
with fitting parameter $p = 0.29$ and $f(M_0) = 0.52$. Our choice of $M_0$ corresponds to a halo that reaches $T_\Hal{}$ at redshift $z=10$, the latest epoch for \acro{DCBH} candidates to form without overlapping with the universe's reionization period (see e.g.\ the Planck results~\cite{Planck:2016mks}), while still allowing time for these seeds to grow into the supermassive black holes observed at $z\sim 6$.

The halo is an evolving background that affects the gas dynamics. We focus on tracking these dynamics as a one-zone model in the halo core where the gas is effectively isotropic and homogeneous. 
We model the core gas density and temperature by patching together the intergalactic medium and virialized values at the virialization redshift, $z_\text{vir}$,
\begin{align}
    n_\text{p}(z)
    &=
    \begin{cases}
        n_{\text{p}0} (1+z)^3 
        & z > z_\text{vir}
        \\ 
        n_\text{core}(z)
        & z \leq z_\text{vir} 
    \end{cases}
    &
    T 
    =
    \begin{cases}
    T_\text{IGM}(z)
    & z > z_\text{vir}         
    \\
    T_\text{vir}(M,z)
    & z \leq z_\text{vir} 
    \end{cases}    
    \ .
    \label{eq:halo:gas:temp:patch}
\end{align}
The core is a constant density region that is a tenth of the virial radius, $R_\text{core} \sim 0.1\, R_\text{vir}$ as shown by simulations with no cooling mechanisms~\cite{Visbal:2014nogo, Nebrin:2023yzm, Hegde:2023wxz}.
We use the filtering mass formalism to determine the redshift of virialization, $z_\text{vir}$.
This is similar to the Jeans mass but accounts for the halo history~\cite{Gnedin:1997td,Gnedin:2025opp}.
We implement this
with the fitting function in Ref.~\cite[eq.~8]{Hegde:2023wxz} in
limit of zero relative dark matter--baryon velocity,
\begin{align}
    M(z_\text{vir}) &= M_\text{F}(z_\text{vir})
    &
    M_\text{F}(z) 
    &\approx
    1.66 \times 10^4  \, M_\odot 
    \left( \frac{ 1+z }{ 21 } \right)^{0.85} \ . 
    \label{eq:halo:model}
\end{align}
We plot $n_\text{p}(z)$ and $T(z)$ in Fig.~\ref{fig:background:halo:evolution} and present the analytic forms in Appendix \ref{sec:AP:standard:halo:dynamics}.

\begin{figure}[tb]
    \centering
    \includegraphics[width=\linewidth]{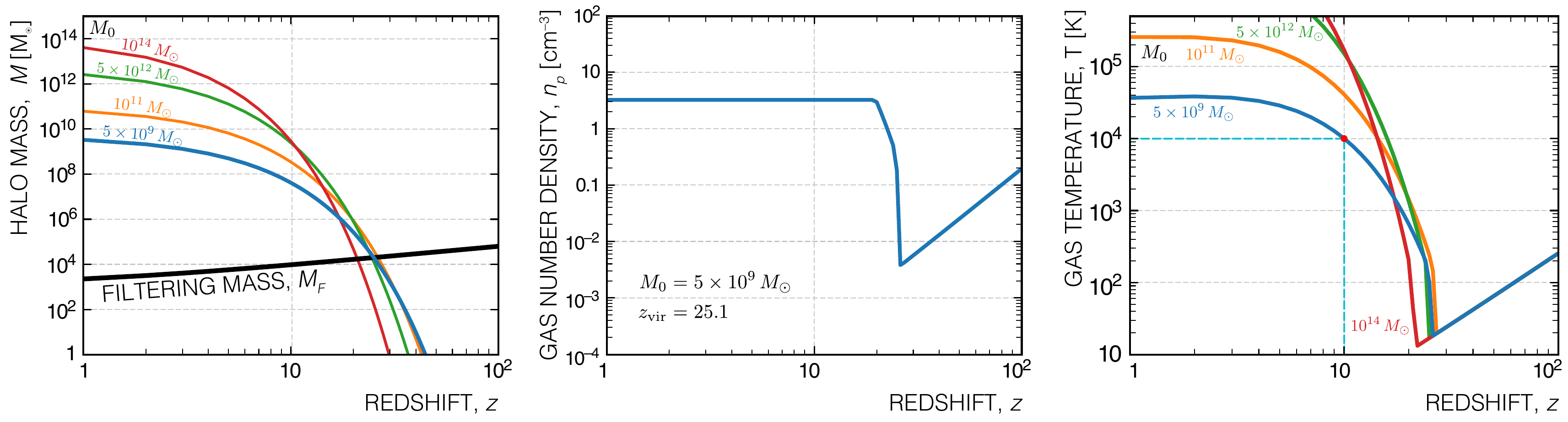}
    \caption{%
        Halo evolution with redshift. 
        \textsc{left}: total halo mass $M(z)$ compared to the filtering mass $M_\text{F}$.
        \textsc{middle}: gas (H nuclei) number density in the core, $n_\text{p}(z)$.
        \textsc{right}: gas temperatures, $T(z)$.
        }
    \label{fig:background:halo:evolution}
\end{figure}

\subsection{Timescales for temperature change}
\label{sec:timescales}
The heating or cooling timescale of a gas of temperature $T$ is $\tau \equiv T/\dot{T}$, where $\tau>0$ corresponds to heating and $\tau<0$ corresponds to cooling. When multiple processes contribute to the temperature change, one may express the total rate of change of the gas temperature as
\begin{align}
    \frac{ \D{T} }{ \D{t}  }
    &=
    \frac{2}{3 k_\text{B} \, n_\text{p} }
    \left(\Gamma - \Lambda\right)
    &
    \frac{ \D{\,\ln T} }{ \D{t}  }
    &\equiv 
    \sum_i \frac{1}{\tau_i}
    \ ,
    \label{eq:dT:dt:general:and:tau:def}
\end{align}
where $\Gamma = \sum_a \Gamma_a$ is the heating rate from the set of processes $\{a\}$ that increase the temperature, and $\Lambda = \sum_b \Lambda_b$ is the cooling rate from the set of processes $\{b\}$ that decrease the temperature. In the second equation we write this in terms of the inverse temperature change timescales for each process $i \in \{a\}\cup \{b\}$. 
The normalization on the left-side comes from approximating the internal energy density at low temperatures as that of a monatomic gas: 
$u = \tfrac{3}{2}n k_\text{B} T$, though other normalizations are used in the literature.\footnote{For example, Gnedin and Hollon write $\dot U = n_\text{p}^2(\Gamma - \Lambda)$ so that $\Gamma$ and $\Lambda$ are density independent in the limit of collisional ionization equilibrium~\cite{Gnedin:2012qk}.}
The quantity $n_\text{p} = n_{\Hm{}} + n_{\Ha{}} + 2n_{\HH{}} + \cdots$ is the number density of H nuclei and $k_\text{B}$ is the Boltzmann constant, which we leave explicit to ease comparison between astrophysical temperatures and photon energies.

The temperature-changing processes are \HH{} cooling, \Hal{} (atomic line) cooling, and dynamical heating. The rates for these processes are
\begin{align}
    \Lambda_{\HH{}} 
    &= 
    \lambda_{\HH{}} n_\text{p}^2 x_{\HH{}}
    &
    \Lambda_{\Hal{}} 
    &=
    \lambda_{\Hal{}} n_\text{p}^2 x_\text{e}
    &
    \Gamma_\text{dyn}
    &=
    \frac{3}{2}
    n_\text{p} 
    k_\text{B} 
    \frac{ \D{ T_\text{vir} } }{ \D{t} }
    \ .
    \label{eq:t.heat:def}
    \\
    \intertext{The dynamical heating term ensures that the temperature tracks the virial temperature when cooling is subdominant. The associated time scales for these processes are}
    \tau_\HH{}
    & = 
    \frac{3}{2}
    \frac{ k_\text{B} T }{ n_\text{p} }
    \frac{ 1 }{ \lambda_\HH{} x_\HH{} }
    &
    \tau_\Hal{} 
    & = 
    \frac{3}{2}
    \frac{ k_\text{B} T }{ n_\text{p} }
    \frac{ 1 }{ \lambda_\Hal{} x_\text{e} }
    &
    \tau_\text{dyn} 
    & = 
    \frac{T}{ \D{ T_\text{vir} } \!/\! \D{t} }
    \ . 
    \label{eq:t.heat}
\end{align}
The fraction $x_i = n_i/n_\text{p}$ is the ratio of the number densities between species $i$ and H nuclei, see footnote~\ref{foot:xi:ratio}.
All quantities other than the Boltzmann constant $k_\text{B}$ depend on redshift, if implicitly through their dependence on the gas temperature $T$.

The \HH{} cooling rate coefficient is fit in the low density limit by~\cite[eq.~A.7]{Galli:1998dh},
\begin{align} 
    \log_{10}
    \frac{ 
        \lambda_{\HH}(T)
        }{
        \text{erg} \, \text{cm}^3 \, \text{s}^{-1}
        }
    & = 
    - 103 
    + 97.6\, \ell_T 
    - 48.1\, \ell_T^2 
    + 10.8\, \ell_T^3  
    - 0.903\, \ell_T^4
    \
    \ ,
\end{align}
where\footnote{Our definition of $\ell_T$ corrects the dimensionally inconsistent $\log T_\text{g}$ in Ref.~\cite{Galli:1998dh}.} $\ell_T \equiv \log_{10}({T}/{\text{K}})$.
The \Ha{} cooling rate coefficient is fit by~\cite[eq.~15a]{1992ApJS...78..341C}
\begin{align}
    \lambda_{\Hal}
    & =
    7.5 \times 10^{-19}\, \text{erg}\, \text{cm}^3 \, \text{s}^{-1}
    \;
    \left[
        1 
        + 
        \left( \frac{T}{10^5 \, \text{K}} \right)^{1/2} 
    \right]\inv
    \exp\!{ \left(\frac{- 118348 \ \text{K}}{T} \right)}
    \ .
    \label{eq:lambda:Hal}
\end{align}

\begin{figure}[t]
    \centering
    \includegraphics[width=\linewidth]{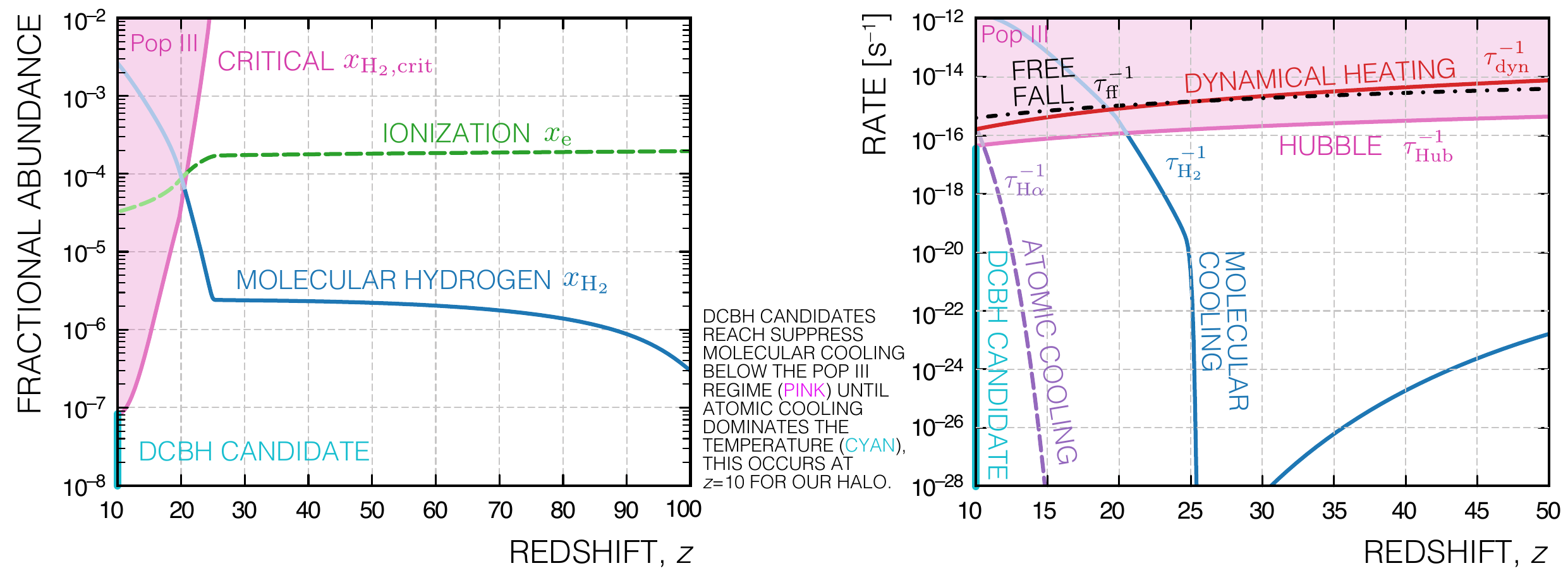}
    \caption{
    \textsc{left}:
    Model halo history in the absence of additional photons from axion decay. 
    \popiii{} star formation begins when the \HH{} fraction, $x_\HH{}$, crosses the critical fraction. 
    The halo properties are constructed so that in the absence of \HH{}, atomic cooling would begin at the left edge: $z=10$ ($T=10^4\,\text{K}$). 
    The $x_\text{e}$, critical \HH{}, baryon density curves do not change when photons from axion decay are introduced.
    \quad
    \textsc{right}: The heating (red), \HH{} cooling (solid blue), \Hal{} cooling (dashed purple), free fall rate (dashdot black), and Hubble rate (pink) in this work. The conservative condition for forming a heavy seed candidate is that the \HH{} cooling rate stays below the Hubble rate until the atomic cooling threshold ($T=10^4\text{K}$, left edge) when the \Hal{} cooling activates.
    The condition that \HH{} cooling is slower than the free fall rate is a weaker, but more realistic condition. We choose the conservative condition because our halo model breaks down when the \HH{} cooling rate surpasses the dynamical heating rate (red); at this point the halo temperature no longer tracks the virial temperature and a more detailed analysis is necessary.
    }
    \label{fig:standard:halo:history:and:thermal.rates}
\end{figure}

\paragraph{Halo model}
The benchmark halo \eqref{eq:MAH} is conservative because halos that reach atomic cooling at larger redshifts would have done so with a relaxed set of conditions because they would have
(1) a larger photon flux from the \acro{IGM} axion decay,
(2) more dynamical heating 
(3) a shorter characteristic halo timescale against which the cooling timescale is compared.
Fig.~\ref{fig:standard:halo:history:and:thermal.rates} (left) plots the model halo history in the absence of dark matter decays; we refer to Appendix~\ref{sec:halo:time:evolution} for details of this history. 
As the halo evolves in time, molecular hydrogen builds up and eventually surpasses the critical fraction $x_{\HH{},\text{crit}}$ (defined below). When this occurs, we assume that the halo goes on to form \popiii{} stars rather than \acro{DCBH} candidates.
Additional photons from axion decay decrease the \HH{} fraction, $x_\HH{}$, but otherwise do not directly affect the other quantities.

\paragraph{Halo timescale and model validity}
In \eqref{eq:tauff_rhobar} present the free fall time, $\tau_\text{ff}$, as a natural choice for the halo timescale in determining the critical \HH{} fraction, $x_{\HH{},\text{crit}}$, see Ref.~\cite[App.~A]{Omukai:2000ic}. However, we use a more conservative quantity%
, the Hubble time, $\tauHub{}(z)$, which is an order of magnitude longer than $\tau_\text{ff}$ e.g.~\cite{Tegmark:1996yt}.  This choice forces the halo evolution to remain within the regime of validity of our halo model which assumes that the halo temperature tracks the virial temperature. This occurs when dynamical heating is the dominant temperature changing mechanism, see \eqref{eq:t.heat:def} and \eqref{eq:dT:dt:general:and:tau:def}.

The model breaks down when the \HH{} cooling rate approaches the dynamical heating rate. Such a scenario is shown in Fig.~\ref{fig:standard:halo:history:and:thermal.rates} (right) for the benchmark halo with no photons from axion decay. When the cooling rates are significant, the gas no longer tracks the virial temperature. The figure shows that there is a small window where the cooling rate is below the free-fall rate $\tau_\text{ff}$, but above the dynamical heating rate $\tau_\text{dyn.}$.

We thus take as a conservative condition the requirement for \acro{DCBH} candidacy that the \HH{} cooling timescale is longer than the Hubble time scale. This defines the critical \HH{} fraction in \eqref{eq:win:condition}, see Ref.~\cite[App.~A]{Omukai:2000ic}:
\begin{align}
    \text{\textbf{DCBH candidate:}}
    &
    &
    \tau_{\HH{}}
    &> 
    \tauHub{}
    &
    & \Longleftrightarrow &
    x_\HH{}
    <
    x_{ \HH{}, \text{crit} }
    & \equiv
    \frac{
        3 k_\text{B} T
        }{
        2 \lambda_\HH{} n_\text{p} \tauHub{}
        } 
    \ ,
    \label{eq:.xH2.crit}
\end{align}
where all quantities are $z$-dependent other than the Boltzmann constant $k_\text{B}$. A \acro{DCBH} candidate satisfies this condition until the gas reaches the atomic cooling limit, which is set to $z=10$ in our benchmark halo.

\subsection{Dynamics}

We solve the chemical dynamics of the gas for the \HH{} fraction $x_\HH{}$ in the presence of photons from axion decay. We define \acro{DCBH} candidates are those for which $x_\HH{}$ satisfies \eqref{eq:win:condition} with $x_{ \HH{}, \text{crit} }$ in \eqref{eq:.xH2.crit}.
We take as our initial condition 
\begin{align}
    x_\HH{}(z_\text{init}) &= 10^{-8}
    & 
    z_\text{init} &= 120
    \ .
    \label{eq:xH2:initial:condition}
\end{align}
and evolve the system to $z=10$, taking into account halo formation and gas collapse in accordance with the standard \acro{$\Lambda$CDM} paradigm. 
We assume that halos that do not satisfy \eqref{eq:win:condition} 
undergo \popiii{} star formation---perhaps with a delayed onset\footnote{The astrophysical signatures of this delayed onset are outside the scope of this study.}.
We describe the evolution of the halo chemistry in the following paragraphs.

\paragraph{Rate coefficient nomenclature} 
We label the rate coefficients for collisional processes as $C_i$, with $i$ denoting the primary end product in a rate: $\dot{x}_i = C_i x_j x_k$ for the process $j+k \to i + (\text{other})$. For example, $C_\Hm{}$ and $C_\HH{}$ are rate coefficients with dimensions of volume per time for the processes in \eqref{eq:h2:hm:form}. We label the rate coefficients of photonic processes as $k_i$, which are calculated according to 
\begin{align}
     k_i = 
     4\pi 
     \int_0^\infty 
     \D{\nu} \;
     \frac{J(\nu)}{h\nu} \,
     \sigma_i(\nu) 
     \ .
     \label{eq:photon.rate}
\end{align}
Here $\sigma_i$ is the interaction cross section and $J (\nu)$ is the mean intensity of photons which has the dimensions of $J_{21}$ given below in \eqref{eq:J21}. We choose not to write $\hbar = 1$ in order to facilitate straightforward dimensional analysis to show that $k_i$ has the dimensions of a rate.

\paragraph{Free electron fraction}
The free electron fraction $x_\text{e}$ is governed by recombination,
$
    \text{H}^+ + \text{e}^{-} \to \text{H} + \gamma 
$,
with a rate equation
\begin{align}
    \Dot{x}_\text{e} &= 
    - C_\text{B} n_\text{p} x_\text{e}^2 \ . 
    \label{eq:H_rate}
\end{align}
We drop a term that accounts for ionizing \acro{CMB} photons that is only relevant at much higher redshifts~\cite[Eq.~24]{Qin:2023kkk}. Recombinations directly to the \Ha{} ground state result in no net ionization since the photon quickly ionizes a nearby \Ha{} atom. 
The case B recombination coefficient $C_\text{B}$ ignores these ground state recombinations. It is fit by~\cite[eq.~14.6]{Draine:2011ism}
\begin{align}
    C_\text{B}
    &=
    2.5\times 10^{13} \,\text{cm}^3\,\text{s}^{-1}
    \left(
        \frac{T}{ 10^4\,\text{K} }
    \right)^{
        - 0.816
        - 0.021 
        \ln\!\frac{T}{ 10^4\,\text{K} }
    } 
    &
    30\,\text{K}< T < 3\times 10^4\,\text{K}
     \ .
\end{align}
We take the initial free electron fraction from the \texttt{CosmoRec} code~\cite{Chluba:2010ca, Chluba:2010fy},
\begin{align}
    x_\text{e}(z_\text{init}) &= 2 \times 10^{-4}  
    &
    z_\text{init} &= 120
    \ .
\end{align}
Axions in the mass range we consider \eqref{eq:flux.1} do not produce ionizing photons and we assume that there are no nearby halos that have already formed stars. We thus have no sources for ionizing photons to drive the reverse process $ \text{H} + \gamma \to \text{H}^+ + \text{e}^{-}$.

\paragraph{Hydrogen anion}
The rate equation for \Hm{} is
\begin{align}
    \Dot{x}_\Hm{} =
    C_\Hm{} x_\text{e} (1 - x_\text{e}) n_\text{p}
    - C_\HH{} x_\Hm{} (1 - x_\text{e}) n_\text{p}
    - k_{\text{pd}} x_\Hm{} 
    \ .
    \label{eq:Hm_rate}
\end{align}
These correspond to the 
processes in \crefrange{eq:h2:hm:form}{eq:hm.dest}. 
We drop a hydrogen ion--anion neutralization term that is negligible for the modest ionization fraction ($x_\text{e} \sim 10^{-4}$) in this environment~\cite[eq.~4]{Qin:2023kkk}. 
The neutral hydrogen number density is $n_\Ha{} \approx (1-x_\text{e})n_\text{p}$.
Because the processes involving \Hm{} are always much faster than the free fall time, we may take the steady state solution of \eqref{eq:Hm_rate} \cite{Hirata:2006bt},
\begin{align}
    x_\Hm{}
    & = 
    \frac{ 
        C_\Hm{} x_\text{e} 
        }{
        C_\HH{} +
        \frac{
            k_\text{pd}
            }{
            (1 - x_\text{e}) n_\text{p}
            }
        } \ .
    \label{eq:Hm_steady}
\end{align}
We present the photodetachment rate in \eqref{eq:photodetachment:xsec}. The rate coefficients are based on Ref.~\cite{1998ApJ...509....1S}, 
\begin{align}
    C_\Hm{} &= 3 \times 10^{-16} \ \text{cm}^3 \ \text{s}^{-1}
    \left(\frac{T}{300 \ \text{K}} \right)^{0.95}
    \exp{\left( -\frac{T}{9320 \ \text{K}} \right)} \ , 
    \\
    C_\HH{} &= 1.5 \times 10^{-9} \ \text{cm}^3 \ \text{s}^{-1}
    \left(\frac{T}{300 \ \text{K}} \right)^{-0.1}
    \ .
\end{align}

\paragraph{Molecular hydrogen}
The production rate for \HH{} from \eqref{eq:h2:hm:form} is
\begin{align}
    \Dot{x}_\HH{}
    & = C_\HH{} x_\Hm{} (1 - x_\text{e}) n_\text{p}
    - x_\HH{} k_\text{LW} 
     \ ,
    \label{eq:HH_rate}
\end{align}
where we may insert the steady state result \eqref{eq:Hm_steady} for $x_\Hm{}$ and photodissociation rate is given in \eqref{eq:LyWer:k:full:expression}.

\subsection{Injection of Photons}
\label{sec:J.igm}

\paragraph{The IGM contribution}
The evolution of the halo chemistry specifically depends on the rates for photodetachment, $k_\text{pd}$ in \eqref{eq:Hm_rate}, and photodissociation, $k_\text{LW}$ in \eqref{eq:HH_rate}. Prior studies examined the role of \emph{in situ} injection of photons from dark matter annihilation or decay inside the halo~\cite{Friedlander:2022ovf,Lu:2024zwa}. However, Ref.~\cite{Qin:2023kkk} observed that the cosmic flux of photons from decay in the intergalactic medium is typically larger than the in situ flux. This has the additional benefits that (1) the axion flux is appreciable even prior to structure formation, and (2) redshifting smears the spectrum so that it does not have to be tuned to sit on one of the narrow Lyman--Werner lines to dissociate \HH{}.

\begin{figure}[t]
    \centering
    \includegraphics[width=\linewidth]{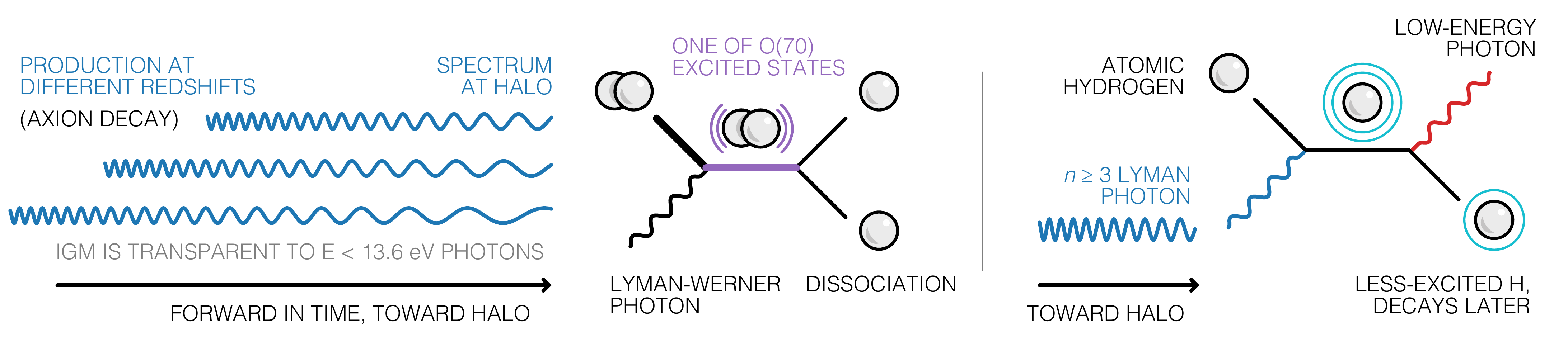}
    \caption{%
    \textsc{Left}: the spectrum of photons from axion decay in the intergalactic medium (IGM) smeared out by reshifting and may excite some of the $\mathcal O(70)$ Lyman--Werner states of \HH{} that could then decay into 2\Ha{}.
    \;
    \textsc{Right}: Photons above $13.6\,\text{eV}$ or with one of the $n \geq 3$ Lyman energies do not reach the halo because the \acro{IGM} is opaque to these photons due to a large density of \Ha{}.
    }
    \label{fig:qualitative}
\end{figure}

\begin{figure}[t]
    \centering
    \includegraphics[width=\linewidth]{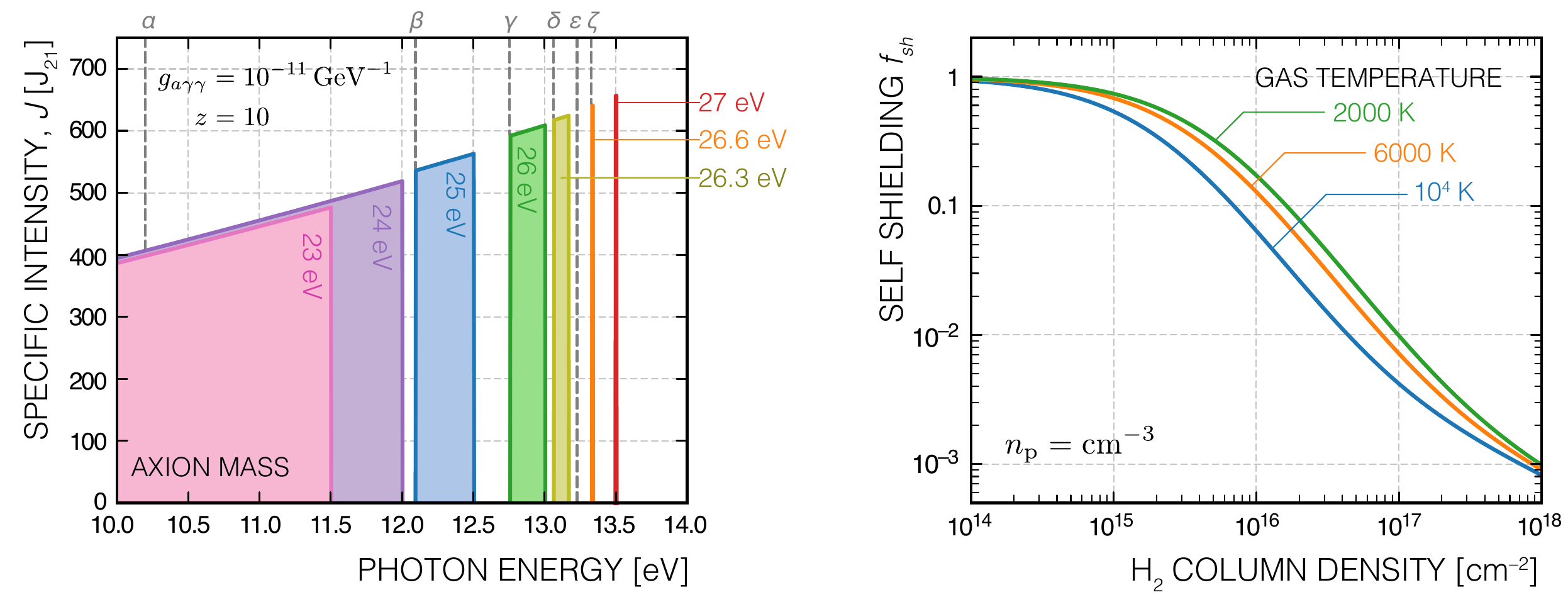}
    \caption{%
    \textsc{Left:} Photon intensity contribution and spectral profile at the halo from axion decay in the intergalactic medium. The axion coupling is set to $g_{a \gamma\gamma} = 10^{-11} \, \text{GeV}^{-1}$ and the observation redshift is $z=10$. 
    For each color, the right boundary is $m_\text{a} / 2$, and the left boundary is the position of a $n\geq 3$ Lyman line for atomic hydrogen. The dashed grey lines show first 6 of these Lyman lines.
    \textsc{Right:} 
    \HH{} self-shielding factor as a function of the \HH{} column density in the conditions of interest, $n_\text{p} = 1 \, \text{cm}^{-3}$. The column density is the product of the core density and the core radius, which is approximately a tenth of the virial radius, $R_\text{core} \sim \mathcal O(10^3\,\text{pc})$ at $z=10$.
    }
    \label{fig.intensity}
\end{figure}

\paragraph{Redshifting and horizons{}}
Redshifting populates photon energies below $m_a/2$. However, the intergalactic medium is opaque to photon frequencies that excite $n \geq 3$ Lyman lines of the \Ha{} atom~\cite{Haiman:1999mn,Ahn:2008uwe}.\footnote{The $n=2$ Lyman excitation is an exception because it re-emits a photon of the same frequency.}
This means that photons from axion decay propagate until they are redshifted to an energy $E^\text{Ly}_i$, the nearest Lyman line below the emission energy $m_a/2$, after which they are effectively absorbed; see Figs.~\ref{fig:qualitative} and~\ref{fig.intensity}.
The gap between these energies gives the maximum emission redshift $z_\text{max}$ from which photons in the range \eqref{eq:flux.1} can propagate to the halo by redshift $z$:
\begin{align}\displaystyle
\frac{ z_\text{max} + 1 }{ z+1  }
= 
\begin{cases}
    \displaystyle
    \frac{ m_{a} }{ 2 E_i^\text{Ly} }
    & 13.6\,\text{eV}
        > \frac{m_a}{2} 
        > 12.1\,\text{eV}
    \\[1ex]
    \infty 
    & 12.1\,\text{eV}
        > \frac{m_a}{2}
        > 0.75\,\text{eV}
\end{cases}
\ .
\label{eq:zmax}
\end{align}
The lowest Lyman excitation energy is $E^\text{Ly}=12.1\,\text{eV}$; photons produced below this energy are not absorbed and so $z_\text{max} = \infty$. 
The screening from Lyman lines becomes more effective for heavier axion masses because the spacing between the Lyman line decreases.

\paragraph{Mean specific intensity}
The rates for photon-initiated processes \eqref{eq:photon.rate} depends on the \emph{mean specific intensity} of photons\footnote{This is a curious quantity that is common in astronomy; we refer to Rybicki and Lightman~\cite[Ch.~1.3]{rybicki2024radiative} for background. A more natural quantity is the radiation energy density $u(\nu, \hat n)$ in a cylinder oriented in the $\hat n$ direction with cross sectional area $\D{A}$ and length $c\,\D{t}$ with a small spread of solid angles $\D{\Omega}$ about $\hat n$. The mean intensity is related by $\tfrac{4\pi}{c} J(\nu) = \int\D{\Omega}\, u(\nu, \hat n)$. } 
from axion decay, $J(\nu)$. 
It is measured in units\footnote{Some papers use an alternative where $J_{21} = J\,/\,[10^{-21} \ \text{ergs} \ \text{s}^{-1} \ \text{Hz}^{-1} \ \text{cm}^{-2} \ \text{str}^{-1}]$ is a dimensionless \emph{measure} of flux, $J$, rather than a constant \emph{unit} of flux. We use a notation amenable to particle physicists.}
\begin{align}
    \Jtwo{}
    = 10^{-21} 
    \, \text{ergs} 
    \; \text{s}^{-1} 
    \, \text{Hz}^{-1} 
    \, \text{cm}^{-2} 
    \, \text{str}^{-1} 
    \ .
    \label{eq:J21}
\end{align}
This encodes the energy (ergs) passing through each cross sectional area (cm$^{-2}$) per unit time (s$^{-1}$) in a given frequency band (Hz$^{-1}$) and coming from a specific solid angle (str$^{-1}$).

The intergalactic medium mean photon intensity of frequency $\nu$ observed at redshift $z$ is 
\begin{align}
    J_\text{IGM}(\nu; z)
    &=
    \frac{ 1 }{ 4 \pi }
    \int_{z}^{z_\text{max}}
    \frac{ \D{z'} }{ (1+z')\, H(z')}
    \;
    \frac{ \D{ N_\gamma } }{ \D{ \nu' } }
    \;
    h\nu'
    \;
    \Gamma_a
    \;
    n_a(z')
    \frac{ \D{V_{z'}} }{ \D{V_z} } \ .
    \label{eq:flux:intermediate}
\end{align}
The integration variable $z'$ is the emission redshift with a maximum value $z_\text{max}$ from \eqref{eq:zmax}.
The fraction with $\D{z'}$ is the proper distance measure $\D{r_\text{p}}(z)$ over a line of sight in the Friedmann--Robertson--Walker metric; we set the speed of light to $c=1$. 
$H(z) \approx H_0 \Omega_{\text{m}0}^{1/2} (1+z)^{3/2}$ is the Hubble rate in the redshift range of interest, with $H_0 = 67.4 \, \text{km} /\text{s}/\text{Mpc}$ and $\Omega_{\text{m}0} = 0.313$ in accordance with Planck 2018 release~\cite{Planck:2018vyg}.
The emission frequency $\nu'$ is a function of the emission redshift $z'$, the observation frequency $\nu$, and the observation redshfit $z$.
Further, the emission spectrum is monochromatic \eqref{eq:flux.2} with emission frequency $\nu_\text{e}$ so that:
\begin{align}
    \nu'
    &= 
    \frac{1+z'}{1+z\phantom{'}}
    \nu
    &
    \frac{ \D{N_\gamma} }{ \D{\nu'_z} } &= 2 \delta(\nu' - \nu_\text{e})
    &
    \nu_\text{e} &= \frac{m_a}{2h}
    \label{eq:nuprime:rzzp:axion:spectrum}
    \, .
\end{align}
The axion number density $n_a(z') = \D{ N_a(z') }\!/{ \D{V_{z'}} } = \rho_a(z')\,m_a\inv$ comes with a factor of the volume element rescaling,
\begin{align}
    \frac{ \D{V_{z'}} }{ \D{V_z} }
    =
    \left(
    \frac{ 1 + z\phantom{'} }{ 1+z' }
    \right)^3
    \ ,
\end{align}
in order that the densities are defined relative to the observation redshift, $z$.
We integrate the emission redshifts $\D{z'}$  over the $\delta$-function in \eqref{eq:nuprime:rzzp:axion:spectrum} using the facts that $\nu' = \nu'(\nu, z,z')$ and 
\begin{align}
    \delta(\nu' - \nu_\text{e})
    =
    \frac{1+z}{\nu}
    \delta\!\left(
        (1+z') - 
        (1+z)
        \frac{\nu_\text{e}}{\nu}
        \right) \ .
        \label{eq:J:delta:function:in:zp}
\end{align}
Whether or not the $\delta$-function has support between $z$ and $z_\text{max}$ depends on the value of $z_\text{max}$ in \eqref{eq:zmax}. Assuming there is support, the $\delta$-function fixes the integration variable to be $z' = z_\text{e}(\nu,\nu_\text{e},z)$, the redshift at which a photon of frequency $\nu_\text{e}$ is redshifted to $\nu$ at redshift $z$,
\begin{align}
    \frac{1+z_\text{e}}{1+z}
    & = 
    \frac{\nu_\text{e}}{\nu}
     = \frac{m_a}{2E_\nu} 
    &
    E_\nu &\equiv h\nu
    \ .
\end{align}
Since these dynamics occur in the matter dominated era and because we are specifically considering the intergalactic axion distribution given in~\cref{eq:axion:density}, the Hubble rate and number densities satisfy
\begin{align}
    H(z_\text{e})
    &= 
    H(z)
    \left(
        \frac{1+z_\text{e}}{1+z}
    \right)^{\!3/2}
    &
    n_a(z_\text{e})
    &= 
    \frac{\rho_a(z)}{m_a}
    \left(
        \frac{1+z_\text{e}}{1+z}
    \right)^{\!3} \ .
\end{align}
We insert these factors into the $J_\text{IGM}(\nu;z)$ expression \eqref{eq:flux:intermediate} to find
\begin{align}
    J_\text{IGM}(\nu;z) = 
    \begin{cases}
        \displaystyle
        \frac{1}{4\pi} 
        \frac{2h\Gamma_a}{H(z)}
        \frac{\rho_a(z)}{m_a}
        \left(\frac{E_\nu}{m_a/2}\right)^{\!3/2}      
        &
        \text{if $z_\text{e}\leq z_\text{max}$}
        \\[2ex]
        0 & \text{otherwise}
    \end{cases}
    \ ,
    \label{eq:flux.4} 
\end{align}
where the condition amounts to assuring that the $\delta$-function in \eqref{eq:J:delta:function:in:zp} has support in the integration range set by the Lyman lines in \eqref{eq:zmax}.
The condition $z_\text{e}\leq z_\text{max}$ is always satisfied for observed energies and axion masses such that $E_\nu \leq m_a/2 < 12.1\,\text{eV}$, because there are no absorption lines. 
For axion masses satisfying $12.1\,\text{eV} < m_a/2 < 13.6\,\text{eV}$, the condition is satisfied for $E^\text{Ly}_i < E_\nu \leq m_a/2$, where $E^\text{Ly}_i$ is the closest $n\geq 3$ Lyman line below $m_a/2$.
Fig.~\ref{fig.intensity} (left) shows the spectra shape for various axion masses. We compare the relative intensity of the intergalactic medium contribution to the \emph{in situ} halo contribution in Appendix~\ref{sec:local:contribution}.

\subsection{Rate Coefficients}

The photons from axion decay contribute to the photodetachment of hydrogen anion \eqref{eq:Hm_rate} and to the photodissociation of molecular hydrogen \eqref{eq:HH_rate}. The rates for these processes follows from inserting the photon flux \eqref{eq:flux.4} into the rate coefficient expression \eqref{eq:photon.rate}.

\paragraph{Photodetachment}  The photodetachment cross section for photons with frequency larger than a threshold frequency $\nu_\text{th}$ is\footnote{%
    The expressions for $\sigma_\text{pd}$ in the literature leave dimensionful factors implicit.  Ref.~\cite[Tab.~4, (23)]{Abel:1996kh} is missing a factor of $\text{Hz}^{3/2}$ and Ref.~\cite[Tab.~3, (27)]{1987ApJ...318...32S} is missing a factor of $\text{cm}^2\;\text{Hz}^{3/2}$.
    These stem from an earlier study that uses atomic units where the Bohr radius is one~\cite{1972A&A....20..263D}.
    The conversion is explained in Ref.~\cite[App.~B.b]{1987ApJ...318...32S}, which relates the energy to the dimensionless wave number $k$ by $h\nu = (E_0 k^2 + 0.754)\,\text{eV}$, where $E_0 = 13.6\,\text{eV}$ is the ground state \Ha{} ionization energy and is a conversion factor from atomic to \acro{SI} units.
    The final expression $\sigma_\text{pd}(\nu)$ drops a $\mathcal O(10^{-3})$ factor in the denominator from the electron affinity of hydrogen. 
    This footnote is why we keep factors of $h$ and $k_\text{B}$ explicit.
}~\cite{1972A&A....20..263D, 1987ApJ...318...32S,Abel:1996kh}
\begin{align}
    \sigma_\text{pd}(\nu) &= 
    \left( 7.93 \times 10^5~\text{cm}^2 \right) 
    \left[
        \frac{ (\nu - \nu_\text{th}) \; \text{Hz}}{ \nu^2 } 
    \right]^{3/2}
    &
    \text{for }
    h\nu 
    &\geq h\nu_\text{th} 
    = 
    0.76
    \, \text{eV} 
    \ . 
    \label{eq:photodetachment:xsec}
\end{align}

\paragraph{Photodissociation}
The photodissociation rate from Lyman--Werner photons is more involved because it depends on the spectrum of molecular ground and excited states that we label with $\alpha$ and $\dot{\alpha}$ respectively,%
\footnote{In molecular chemistry the vibrational and rotational states are labeled by two indices, $(v,J)$, which we condense into a single index for brevity. The electronic quantum number is denoted with capital Latin letters: $\text{B}$ (Lyman), $\text{C}$ (Werner), and $\text{X}$ (ground state).  In our notation, $\alpha \in \text{X}(v, J)$ and $\dot{\alpha} \in \left\{\text{B}(v',J'), \text{C}(v'',J'')\right\}$.} 
\begin{align}
    k_\text{LW}
    &=
    4 \pi f_\text{sh}
    \sum_{ \alpha \dot{\alpha} }
    \int_{ \nu_{th} }^\infty \D \nu 
    \; 
    \frac{ J(\nu)
        }{ h \nu }  \,
    \sigma_{ \alpha \dot{\alpha} }(\nu)\, 
    \text{P}^\text{dis}_{ \dot{\alpha} }\,
    g_\alpha 
    & 
    \text{for } 
    h\nu \geq 
    h \nu_\text{th} 
    = 
    11.2 \, \text{eV} \ . 
    \label{eq:LyWer:k:full:expression}
\end{align}
Here, $\sigma_{ \alpha \dot{\alpha} }(\nu)$ is the cross section for a Lyman--Werner photon of frequency $\nu$ to excite an \HH{} molecule from state $\alpha$ to $\dot{\alpha}$; 
$\text{P}^\text{dis}_{\dot{\alpha}}$ is the probability that $\dot{\alpha}$ dissociates the molecule;
and $g_\alpha$ is the occupation fraction of \HH{} in the ground state $\alpha$.
$f_\text{sh}$ is a fit for the self-shielding factor that accounts for photon absorption by a column density of \HH{} molecules, $N_{\HH{}} = 0.1 \, x_{\HH{}} n_\text{p} R_\text{vir}$~\cite[eq.~7]{Wolcott-Green:2018hyx}.
We plot $f_\text{sh} (N_{\HH{}}, T)$ in Fig.~\ref{fig.intensity} (right) for $n_\text{p} = 1 \, \text{cm}^{-3}$ with temperatures in the range of this study; we present the functional form in Appendix~\ref{app:self:shielding}.

To further simplify \eqref{eq:LyWer:k:full:expression}, we write each Lyman and Werner line absorption cross section in terms of oscillator strength $f_{\alpha \dot{\alpha}}$~\cite[eq.~6.24]{Draine:2011ism}, 
\begin{align}
    \sigma_{\alpha \dot{\alpha} } (\nu) 
    &= 
    \frac{ \pi e_\text{CGS}^2 }{ m_e c }  
    f_{\alpha \dot{\alpha}} 
    \phi_{ \alpha \dot{\alpha} }(\nu) 
    = \frac{e^2}{4 m_e}
    f_{\alpha \dot{\alpha}} 
    \phi_{ \alpha \dot{\alpha} }(\nu) 
    \ ,
    \label{eq:LW:xsec:al:aldot} 
\end{align}
where we $e_\text{CGS} = 4.8\times 10^{-10} \, \text{esu}$ is the \acro{CGS} value of the electron charge used in the astronomy literature and $e=\sqrt{4\pi/137}$ is the value in natural units used in this paper.
$\phi(\nu)$ is the line profile that satisfies identity $\int_0^\infty \D\nu \phi(\nu) = 1$. The Doppler broadening in each band is small enough that the line profile can be approximated by a Dirac $\delta$-function,
\begin{align}
    \phi_{\alpha \dot{\alpha}} (\nu) = 
    \delta(\nu - \nu_{ \dot{\alpha}  \alpha }) \ . 
\end{align}
With these changes, \eqref{eq:LyWer:k:full:expression} simplifies to 
\begin{align}
    k_\text{LW} 
    & = 
    \frac{\pi e^2 f_\text{sh}}{m_e}
    \sum_{\dot{\alpha} \alpha}
    \frac{J(\nu_{\dot{\alpha} \alpha})}{h \nu_{\dot{\alpha} \alpha}}
    f_{\alpha \dot{\alpha}} 
    \text{P}_{\dot{\alpha}}^\text{dis}
    g_\alpha 
    & 
    h \nu_{\dot{\alpha} \alpha} \in
    [\text{Max}(E_i^\text{Ly}, 11.18 \ \text{eV}), m_a / 2]
    \label{eq:LyWer:k:numeric}
\end{align}
In low density environments, $n_\text{p} \sim \mathcal{O}(1 \, \text{cm}^{-3})$, 
\HH{} typically occupies the ortho-$\text{X}(0,1)$ and para-$\text{X}(0,0)$ ground states with a relative abundance ratio of 3:1, 
such that $g_\text{X(0,1)} = 3/4$ and $g_\text{X(0,0)} = 1/4$~\cite[Fig.~2]{Wolcott-Green:2018hyx}. Details of all available Lyman--Werner transitions below the Lyman limit---and corresponding factors of $f_{\alpha \dot{\alpha} } \times \text{P}_{\dot{\alpha}}^\text{dis}$---are presented in the Appendix \ref{AP:Lyman:Werner:Bands}, see also Ref.~\cite[App.]{Haiman:1999mn}.

\subsection{Numerical Solution}

We solve the combined system of ordinary differential equations for recombination \eqref{eq:H_rate}, and molecular hydrogen formation \eqref{eq:HH_rate} with the steady state solution for hydrogen anions \eqref{eq:Hm_steady}, from $z=120$ to $z=10$. We assume an incident flux from axion decay \eqref{eq:flux.4} which feeds into the rate coefficients for photodetachment 
\eqref{eq:photon.rate} and \eqref{eq:photodetachment:xsec} and photodissociation \eqref{eq:LyWer:k:numeric}.
We solve the initial value problem with the \texttt{solve\_ivp} function in the \texttt{scipy.integrate} package. 
The differential equation for the \HH{} fraction is numerically stiff, and therefore we adopt an implicit Runge--Kutta method by passing \texttt{method='Radau'} into the solver.
%

Our parameter space scans combine two linearly spaced axion mass grids: 
    $m_a \in [1 \, \text{eV}, 22.3 \, \text{eV}]$ with $10$ points 
    and 
    $m_a \in [22.4 \, \text{eV}, 27.2 \, \text{eV}]$ with $100$ points. 
The former grid is kept small since the viable parameter space is strongly ruled out for the model halo, whereas the fine spacing in the latter grid allows us to capture details of the Lyman--Werner bands. We sample the photon coupling a logarithmically spaced grid with $50$ points between
$g_{a \gamma \gamma} \in [10^{-14} \, \text{GeV}^{-1}, 10^{-9} \, \text{GeV}^{-1}]$.
For the parameter space plot with low gas density, Fig.~\ref{fig:counterfactual} (right), we sample $50$ points in the low-$m_a$ range.

\section{Results}
\label{sec:results}
The results of our analysis are as follows.

\begin{figure}[tb]
    \centering
    \includegraphics[width=\linewidth]{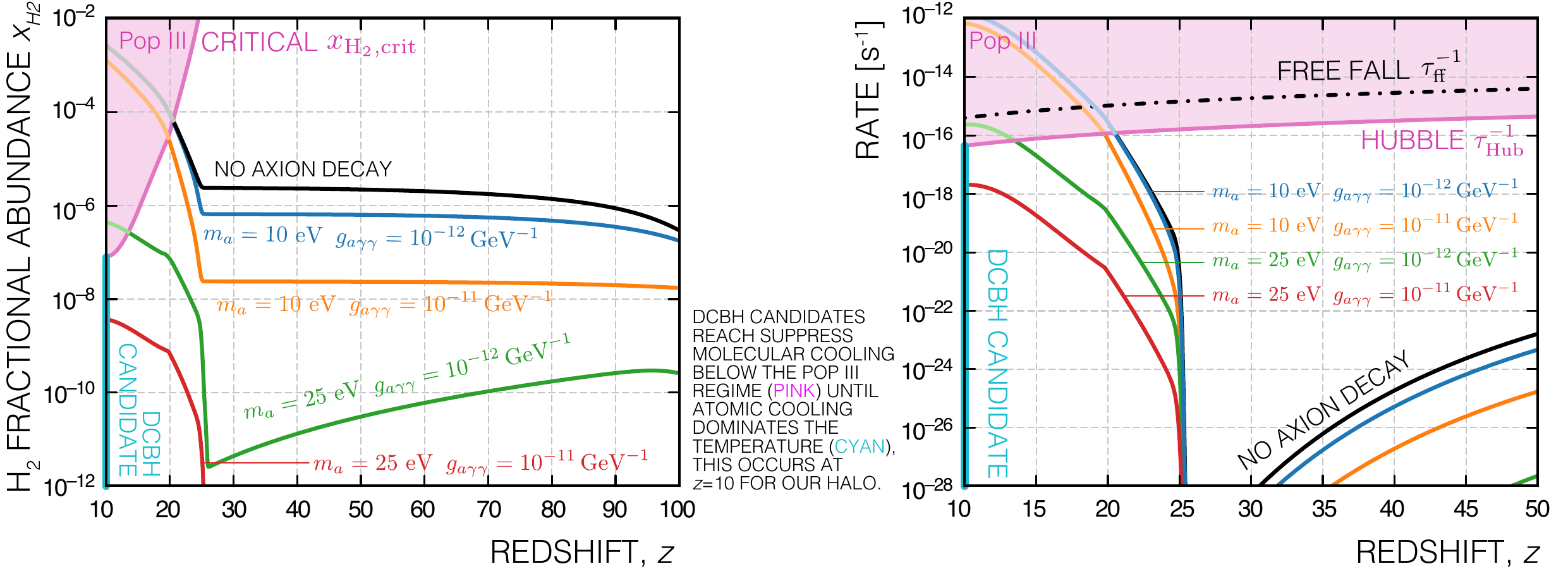}
    \caption{%
        Halo gas evolution with decaying axions: \HH{} fraction (\textsc{left}) and heating/cooling rates (\textsc{right}). Viable \acro{DCBH} candidates remain below red line (\textsc{left}: critical molecular hydrogen fraction, \textsc{right}: Hubble rate) at $z=10$, at which point the halo reaches the atomic cooling limit. This line represents the condition \eqref{eq:.xH2.crit}.
        The benchmark halo follows the history in Fig.~\ref{fig:standard:halo:history:and:thermal.rates}.
    }
\label{fig:histories}
\end{figure}

\subsection{Halo Chemistry with Axion Energy Injection}
\label{sec:sample:halo:evolution}
In Fig.~\ref{fig:histories} we plot the effect of axion energy injection in the intergalactic medium on the evolution of our benchmark halo for a sample of different axion masses and interaction strengths. These plots represent the critical condition \eqref{eq:.xH2.crit} that the \HH{} cooling rate is slower than the Hubble rate. 
Trajectories that remain below this line at $z=10$ for the benchmark halo are \acro{DCBH} candidates.
Trajectories that cross the critical threshold (pink line) produce \popiii{} stars at a delayed redshift compared to the case with no energy injection. 
While this latter population is not the main subject of this work, we point out that our formalism predicts the redshift to which vigorous fragmentation by \HH{} cooling is delayed. These delayed \popiii{} stars may, in turn, be a signature of dark matter energy injection at a level that is weaker than necessary for atomic cooling.

\subsection{Dark Matter Parameter Space}

\begin{figure}[t]
    \centering
    \includegraphics[width=\linewidth]{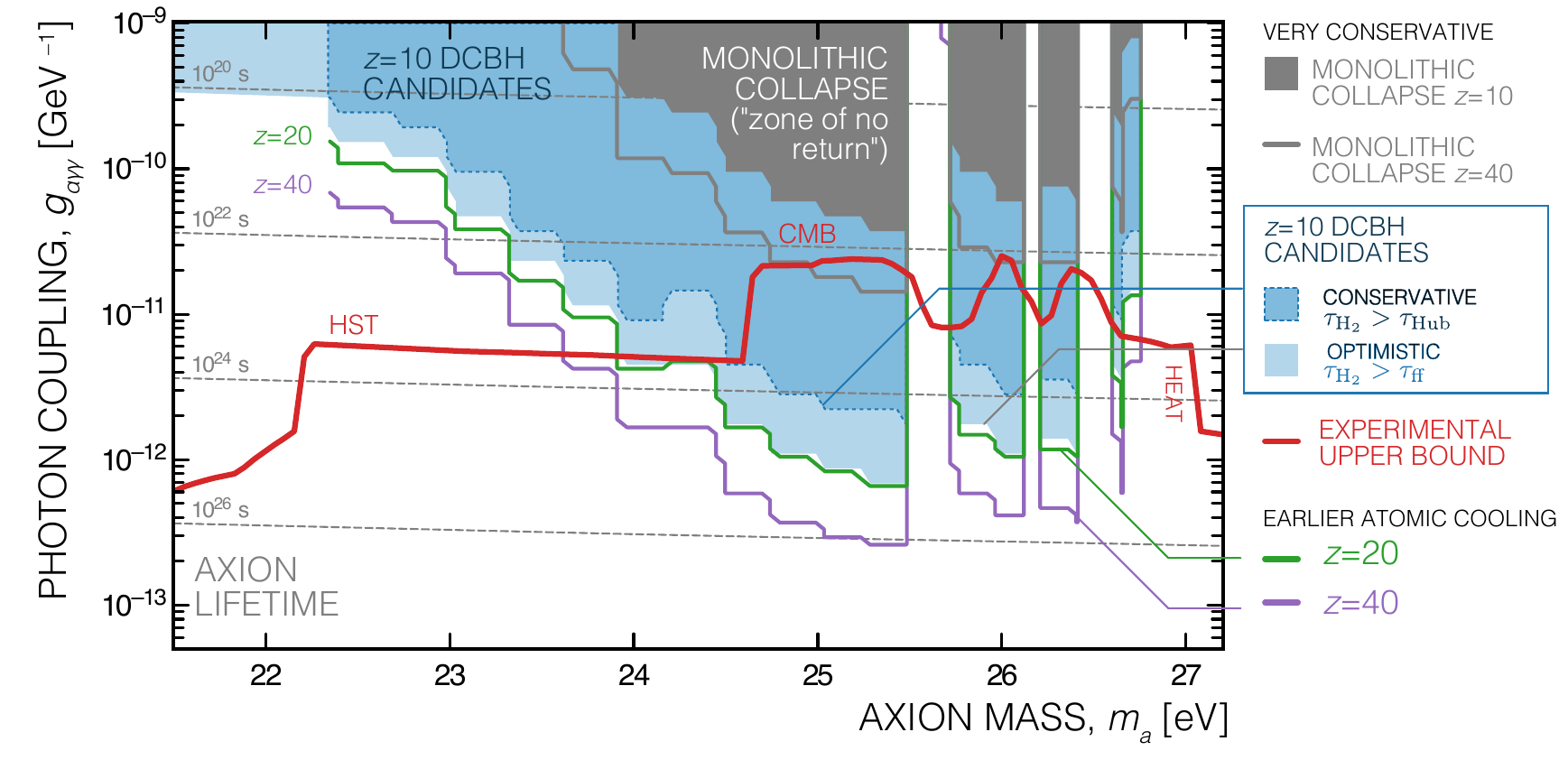}
    \caption{%
    Axion masses and couplings that produce \acro{DCBH} candidates at $z=10$ for conservative (dark blue) and optimistic (light blue) criteria. Also shown are estimated parameters which produces \acro{DCBH} candidates that earlier redshifts (green, purple) and present bounds on axions (red).
    The gray region indicates a Lyman--Werner flux of $J\sim \mathcal{O}(10^3\,\text{J}_{21})$, a benchmark for monolithic collapse~\cite{Wolcott-Green:2011tul}. This is a much stronger condition than our benchmark atomic cooling criteria, see Sec.~\ref{sec:simulations:for:collapse}.
    Constraints on \acro{ALP}s in this range come from measurements of
       dwarf galaxies and galaxy clusters by \acro{HST}~\cite{Todarello:2024qci}, 
        perturbations to the cosmic optical background by \acro{HST} and the New Horizon Space Telescope~\cite{Carenza:2023qxh, Porras-Bedmar:2024uql, Nakayama:2022jza},
        $\gamma$-ray attenuation~\cite{Bernal:2022xyi},
        and perturbations in the cosmic recombination history~\cite{Xu:2024vdn}.
    For axions that decay beyond the Lyman limit, most stringent bounds come from \acro{CMB} distortions~\cite{Bolliet:2020ofj}, and the heating of dwarf galaxies~\cite{Wadekar:2021qae}. 
    These bounds are collected in {AxionLimits}~\cite{AxionLimits}, those in Ref.~\cite{Xu:2024vdn}. 
    }
    \label{fig:Axion.DCBH}
\end{figure}

\noindent
We present the range of axion masses and axion--photon couplings  that can produce \acro{DCBH} candidates in Fig.~\ref{fig:Axion.DCBH}.

\paragraph{Viable region} 
We plot both conservative and optimistic conditions corresponding to the choice of the Hubble time $\tauHub{}$ or the free-fall time $\tau_\text{ff}$ in the critical \HH{} density \eqref{eq:.xH2.crit}.
%
%
The structure of the allowed region is governed by the number of allowed Lyman--Werner lines that are accessible to the photon spectrum, the dissociation strength of these lines, and the screening caused by the \Ha{} Lyman lines---see Fig.~\ref{fig.intensity}. Axions that produce radiation below the Lyman--Werner energy threshold contribute only to photodetachment, and are found to be too ineffective to produce any considerable results in the model halo. 

\paragraph{Monolithic collapse}
The gray region corresponds to a Lyman--Werner flux of $J\sim \mathcal{O}(10^3\,\text{J}_{21})$, which corresponds to a benchmark flux for monolithic collapse into a heavy seed without gas fragmentation~\cite{Wolcott-Green:2016grm}; see our discussion in Section~\ref{sec:simulations:for:collapse}. 
%
We plot this region for ease of comparison to a standard scenario. 

This region is excluded and indicates that the atomic cooling halos realized by the viable parameter space may rely on environmental factors to produce direct collapse black holes, although there is a small region accessible at $ z \gtrsim 40$.


\paragraph{Atomic cooling halos at higher redshift}
Our benchmark halo model is chosen such that a halo with suppressed \HH{} cooling reaches the atomic cooling limit at $z=10$, reflecting the median behavior within the Press--Schechter framework. 
Halos that reach the atomic cooling limit earlier than our benchmark have a relaxed target to form \acro{DCBH} candidates since the photon flux from axion decay scales with redshift as $J_\text{IGM} \propto (1+z)^{3/2}$ in \eqref{eq:flux.4}, and a shorter characteristic time scale raises \eqref{eq:.xH2.crit}. 

To estimate the \acro{DCBH} candidate parameter space for halos that reach atomic cooling at earlier redshifts, we may rescale the results for our benchmark halo. 
We retain chemo-thermal properties of our model at the moment it reaches atomic cooling: 
\begin{align}
    n_\text{p} &= 3.2 \, \text{cm}^{-3}
    &
    T &= 10^4 \, \text{K}
    &
    x_\text{e} &= 3\times 10^{-5} 
    \label{eq:chemothermal:properties}
    \ .
\end{align}
In the presence of an appreciable Lyman--Werner radiation, $J \gtrsim 0.01 \, J_\text{21}$, the processes involving dissociation of \HH{} become much faster than the free fall time.
Furthermore, the axion parameters that produce \acro{DCBH} candidates have a small \HH{} abundance.
Thus, instead of solving the complete \HH{} rate equation in \eqref{eq:HH_rate} as a function of time, we take its steady state solution
\begin{align}
    x_{\HH{},\text{eq}} 
    \simeq 
    \frac{ C_\Hm{} x_\text{e} n_\text{p} }{ k_\text{LW} }
    \left(  1 + 
                \frac{ k_\text{pd} }{ C_\HH{} n_\text{p} }
        \right)\inv 
    \ ,
    \label{eq:H2.equilibrium}
\end{align}
where we may ignore self shielding $f_\text{sh}\approx 1$. This is because successful \acro{DCBH} candidates have 
    a small molecular hydrogen fraction $x_\HH{} < 10^{-7}$ (Fig.~\ref{fig:histories}), 
    a characteristic lengthscale is $R_\text{core}\sim 0.1\,R_\text{vir} \sim \mathcal O(10^3\,\text{pc})$ (see \eqref{eq:virial.radius}),
    and a core density $n_\text{p}\sim \text{cm}^{-3}$ (Fig.~\ref{fig:background:halo:evolution}).
    This gives a \HH{} column density of $10^{-14}\,\text{cm}^{-2}$ and a negligible shielding factor in Fig.~\ref{fig.intensity}.
We show the results of this estimate for \acro{DCBH} candidates that reach atomic cooling at $z=20$ and $z=40$ in Fig.~\ref{fig:Axion.DCBH}.
We also show the monolithic collapse threshold at $z=40$.

\subsection{Critical Curves}
\label{sec:critical:curves}
The traditional way to show the impact of external radiation on heavy seed formation is through \emph{critical curves} that plot the minimum combination of the photodissociation rate $k_\text{LW}$ and photodetachment rate $k_\text{pd}$ to induce direct collapse in a simulation assuming that this flux is constant in time~\cite{Wolcott-Green:2011tul, Wolcott-Green:2016grm}. As explained by Kusenko, Lu, and Picker, these plots do not capture the time evolution of the condition for sufficiently suppressed \HH{} abundance~\cite{Lu:2024zwa}, though they are a convenient way to compare to simulations.

\begin{figure}
    \centering
    \includegraphics[width=\textwidth]{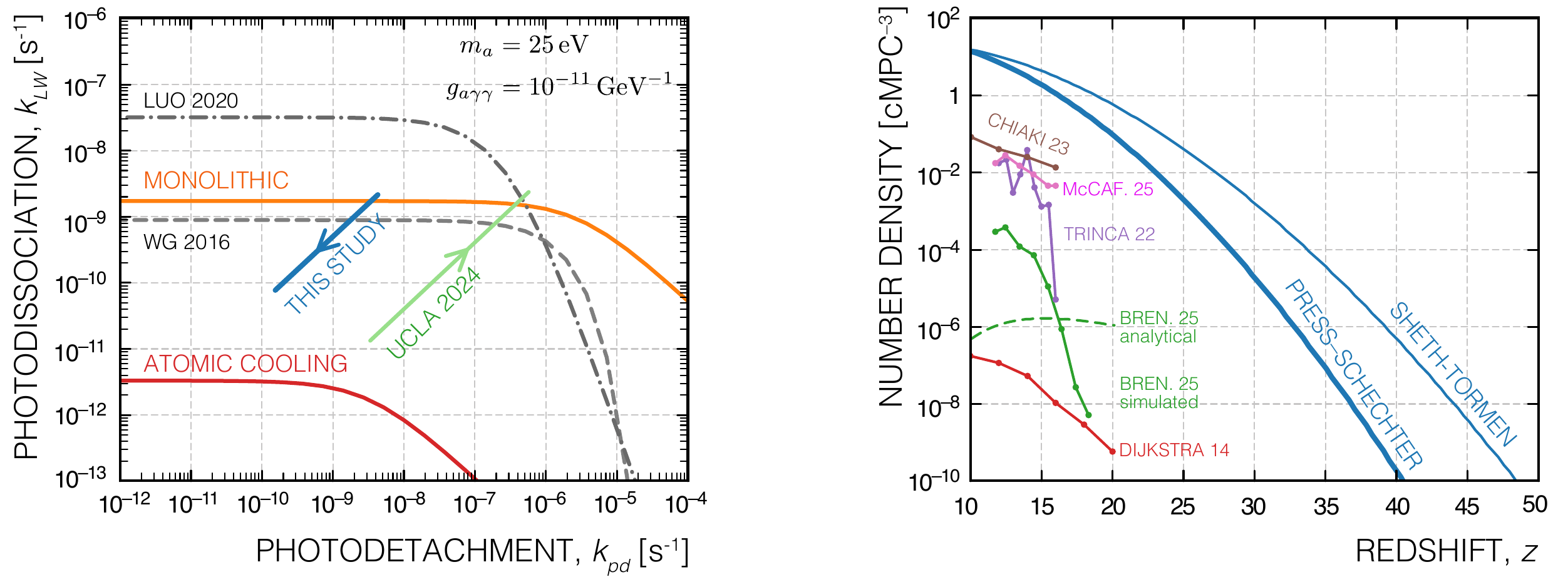}
    \caption{%
    \textsc{left}: Approximate critical curves for monolithic collapse (orange), and atomic cooling halos (red) based on our analysis. The grey lines represent simulated critical curves for monolithic collapse from Refs.~\cite{Wolcott-Green:2016grm} (dashed) and \cite{2020MNRAS.492.4917L} (dot-dashed). The blue line shows the trajectory of an axion model as it evolves in time, from $z=100$ (top right) to $z=10$ (bottom left). 
    The green line shows the trajectory from \emph{in situ} axion decay with adiabatic contraction from Ref.~\cite{Lu:2024zwa}. 
    \quad 
    \textsc{right}: Number density of \acro{DCBH} candidate halos under conditions where our benchmark halo reaches atomic cooling at $z=10$.
    We plot two different mass functions. 
    We compare with heavy seed simulation results from the literature,
    labeled by first authors Chiaki\cite{2023MNRAS.521.2845C}, McCaffrey~\cite{McCaffrey:2024xpt},
    Trinca~\cite{Trinca:2022txs},
    Brennan~\cite{OBrennan:2025sft},
    and Dijkstra~\cite{2014MNRAS.442.2036D};
    these are compiled in Ref.~\cite[Fig.~8]{OBrennan:2025sft}.
    }
    \label{fig:critical.curves}
\end{figure}

In Fig.~\ref{fig:critical.curves}, we present the trajectory of the photon flux from axion decays in the intergalactic medium compared to approximate critical curves and sample critical curves from simulations. Our approximate critical curves are constructed using the equilibrium relation \eqref{eq:H2.equilibrium} with $x_{\HH{},\text{eq}}$ replaced with the critical value $x_{\HH{},\text{crit}}$ and assuming the chemo-thermal properties \eqref{eq:chemothermal:properties}.
Models that lie above these critical curves satisfy $x_\text{\HH{}} < x_\text{\HH{},crit}$. 
We produce an approximate critical curve for monolithic collapse analogously by imposing the chemo-thermal properties in Ref.~\cite[Fig.~7]{Omukai:2000ic},
\begin{align}
    n_\text{p} &= 3 \times 10^3 \, \text{cm}^{-3}
    & T &=8000 \, \text{K}
    & x_\text{e} &= 10^{-4}
    & x_\text{\HH{}} < x_\text{\HH{},crit} = 5\times10^{-7}
    \ . 
\end{align}
This approximate treatment is only meant for ease of comparison to simulations. 
%

\subsection{Estimated Number Density of DCBH Candidates}

When axion decay is able to generate \acro{DCBH} candidates, one would like to compare the distribution of high-redshift supermassive black holes to those that are observed. 
Unfortunately, a direct comparison is challenging because of the mass evolution between the onset of atomic cooling targeted in this study and the observations of high-redshift quasars in the range $7 < z < 10$. 
This evolution may include environmental effects that can accelerate mass accretion and adiabatic contraction that boosts \emph{in situ} halo photons from dark matter decay.

In this study, we compare our estimate of \acro{DCBH} candidate number density over time with predictions from recent simulations, which serve as proxies for distributions that may explain observations of high-redshift quasars. Since our criterion only requires that candidate halos reach the atomic cooling limit, our predicted number density is expected to exceed simulated values.
Assuming that the photon flux realizes the critical condition $\tau_\HH{}>\tauHub$ for our benchmark halo, we estimate the number density of \acro{DCBH} candidates as follows:
\begin{align}
    \left.
    \frac{ \D{ N( z ) } }{ \D{V} } 
    \right|_\text{DCBH} 
    & = 
    \int_{ M_\text{ACH} }^\infty \D{M} 
    \left. 
        \frac{\D N}{\D M\! \D V}(z) 
    \right|_{\tau_{\text{H}_2} > \tauHub}
    \ ,
    \label{eq:ACH.count}
\end{align}
where $\D N\! / \D M\! \D V$ is either 
    the Press--Schechter halo mass function~\cite{1974ApJ...187..425P} calculated using \texttt{HaloMod}~\cite{Murray:2020dcd} 
    or 
    the Sheth--Tormen mass function \cite{1999MNRAS.308..119S} that fits well with \acro{JWST} and simulations at high redshift~\cite{OBrennan:2024jcg}. 
Fig.~\ref{fig:critical.curves} (right)
shows the time evolution of the \acro{DCBH} candidate number density. 
We use the fact that if the axion flux realize atomic cooling in the benchmark halo by $z=10$, then it will necessarily induce atomic cooling in similar halos that reach the atomic cooling threshold temperature at earlier redshifts---see the discussion below \eqref{eq:lambda:Hal}.
For axion parameters that only realize atomic cooling at higher redshifts---say for the $z=20$ estimated line in Fig.~\ref{fig:Axion.DCBH}---one may simply truncate the predicted number density in Fig.~\ref{fig:critical.curves} at $z=20$, followed by a horizontal evolution to lower redshifts.

We see that both mass functions produce a history between one and eight orders of magnitude larger than simulations, with this range reflecting the breadth of results from the simulations themselves.
Unfortunately, without simulations tailored to our scenario and more population data, it is difficult to quantify how well our results fit, beyond noting that they are allowed. 
If, however, the number density were below that of simulations, one would conclude that axion-induced energy injection in the \acro{IGM} likely yields negligible effects.

\subsection{Model Features}

\begin{figure}[tb]
    \centering
    \includegraphics[width=\linewidth]{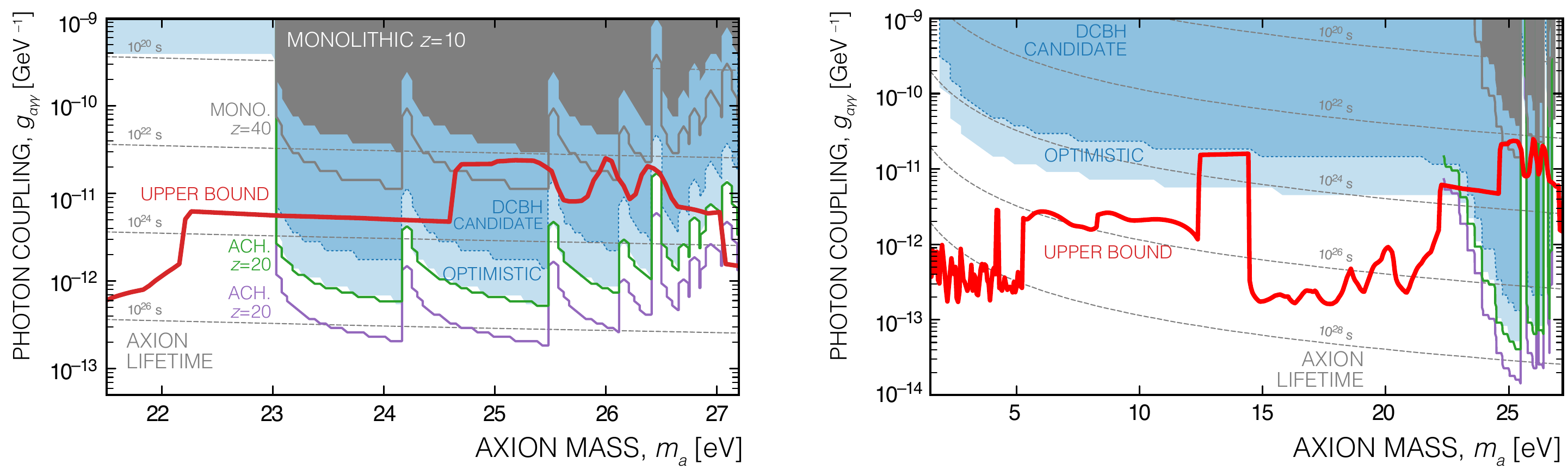}
    \caption{%
        Versions of Fig.~\ref{fig:Axion.DCBH} with different model assumptions.
        \textsc{Left}: Assigning a constant cross section \eqref{eq:sigma:LW:constant} across the Lyman--Werner band. 
        \;
        \textsc{Right}: Reducing the core density by a factor of 30. The upper bound below $m_a<15$\,eV is set by \acro{DESI}~\cite{Wang:2023imi}.
    }
\label{fig:counterfactual}
\end{figure}

In Fig.~\ref{fig:counterfactual} we examine two model modifications to illustrate the significance of the Lyman--Werner lines in \eqref{eq:LyWer:k:numeric} and the sensitivity to the core density.

\paragraph{Constant Lyman--Werner cross section}
The standard treatment of photodissociation is to assume that the photon spectrum is broad compared to the line density so that one may use a constant photodissociation cross section for photons in the Lyman--Werner band in place of \eqref{eq:LW:xsec:al:aldot}; we discuss this approximation in Appendix~\ref{sec:constant:LW:cross:section}. This is a reasonable assumption for blackbody spectra from nearby stars, but breaks down when the spectrum is narrower such as from dark matter decay.  This tracks the `finger' structure in Fig.~\ref{fig:Lyman.Werner.bands}, but spuriously fills in gaps between Lyman--Werner resonances. The most significant change is that our result, Fig.~\ref{fig:Axion.DCBH}, has effectively lost the lowest-energy finger due to the weakness of the Lyman--Werner oscillator strengths below $12\,\text{eV}$.

\paragraph{Gas density}
We find a more drastic difference when reducing the maximum gas density in the core of the halo by an order of magnitude, from $3.2 \, \text{cm}^{-3}$ to $0.1 \, \text{cm}^{-3}$, which is similar to the modeling in Ref.~\cite{Qin:2023kkk}. Gas rarification may occur during from halo--halo\footnote{The Filipino dessert \emph{halo halo} translates to \emph{mix mix}, an apt description of dynamical heating from mergers.} interactions due to dynamical heating that are outside the scope of our treatment; see e.g.~Ref.~\cite{Yoshida:2003rw}.
The reduced gas density suppresses the \HH{} production rate and increases the critical threshold for \popiii{} star formation in \eqref{eq:.xH2.crit}. 
The order of magnitude reduction in $n_\text{p}$ commensurately expands the \acro{DCBH} candidate reach by an order of magnitude in the axion coupling. In Fig.~\ref{fig:counterfactual} (right) we show that this brings the \acro{DCBH} candidate region closer the experimentally viable region in the $13\text{--}15\,\text{eV}$ photodetachment axion mass range.

\section{Conclusion and Discussion}
\label{sec:conclusion}

We semi-analytically study the impact of photon injection from dark matter decay in the intergalactic medium on the formation of atomic cooling halos---that are \emph{direct collapse black hole} (\acro{DCBH}) \emph{candidates}---in the era leading to cosmic dawn.  
The decay spectrum cosmologically redshifts and overlaps with the Lyman and Werner transitions of the molecular hydrogen ground state.
Exciting these lines can deplete the molecular hydrogen abundance enough to prevent gas fragmentation and the formation of \popiii{} stars. Sufficiently irradiated halos would reach the atomic cooling theshold, a necessary condition for forming heavy seed black holes through direct collapse.

Our one-zone model evolves the gas temperature and chemistry in pre-stellar, metal-free halos---the cradles of the first stars. 
We make conservative assumptions for the background halo model, though our treatment does not capture the subsequent evolution from atomic cooling to direct collapse.
A novel feature of our model is the modeling of the Lyman--Werner transitions as individual lines rather than a band with a constant cross section. 
Our axion model captures the generic behavior spin-0 dark matter that decays into photon pairs.

Our condition for a successful \acro{DCBH} candidate is that the \HH{} cooling time is longer than the Hubble time, \eqref{eq:.xH2.crit}, until the gas heats to the atomic cooling temperature $T_\Hal{}$, \eqref{eq:win:condition}. We include the role of dynamical heating by tracking the virial temperature with the mass accretion history formalism and a benchmark halo, \eqref{eq:MAH}.

\subsection{Summary of Results}

\begin{enumerate}
\item  The main result of our study is the allowed region in the axion--photon coupling and axion mass parameter space that admits \acro{DCBH} candidates, Fig.~\ref{fig:Axion.DCBH}: a window of axion masses between 24.5--26.5\,eV with axion--photon couplings as low as $3\times 10^{-12}\,\text{GeV}\inv$. This region has gaps due to the absence of \HH{} transitions, but otherwise extends an order of magnitude below existing bounds on the coupling. 
Compared to dark matter decays in the halo, the intergalactic medium contribution suppresses the \HH{} abundance before the gas virializes. This contributes towards its effectiveness against \HH{} self-shielding.

\item  We find that photodetachment, which suppresses the \Hm{} precursor for \HH{} production, is inefficient for inducing \acro{DCBH} candidates. Unlike photodissociation, the required axion--photon couplings for suppressing the \HH{} abundance by photodetachment is incompatible with axion bounds. 

\item Astrophysical studies of \HH{} treat the Lyman--Werner transitions as a continuous band with a constant cross section. This breaks down when the photon spectrum is narrow compared to the band, as is the case for dark matter decay.
We find that the constant cross section assumption can substantially overestimates the \HH{} dissociation rate. 

\item Our results are sensitive to the core gas density, $n_\text{p}$, since the \HH{} production rate is proportional to this quantity. We find that the minimum in the axion--photon coupling for atomic cooling scales roughly with the core density within our model. 
We present the counterfactuals in this, and the prior point, in Fig.~\ref{fig:counterfactual}, which is to be contrasted with our primary result, Fig.~\ref{fig:Axion.DCBH}.

\item An estimate of the \acro{DCBH} candidate number density based on our halo model is greater than heavy seed density in simulations. This is consistent in that one expects a fraction of our candidates to become heavy seeds. However, more quantitative connections to observations require (1) simulations that determine this fraction, and (2) more data on the supermassive black hole distribution for $z\gtrsim 5$.

\item Early halos that do not form \acro{DCBH} candidates are presumed to produce \popiii{} stars. this population, however, is delayed compared to the case without axion decay, see~Fig.~\ref{fig:histories}. 

\end{enumerate}

\subsection{Uncertainties}

Our semi-analytic one-zone treatment has transparent chemo-thermal evolution, but lacks the details of large simulations. Quantitative uncertainties will likely require simulations dedicated to a photon flux from dark matter decay.

\paragraph{Subsequent halo evolution}
Although current simulations suggest that the onset of atomic cooling is a sufficient condition for \acro{DCBH} formation, the gas evolution beyond this point lies outside the scope of our work.
This limitation arises because not only does the thermal evolution depend on gas density, but also additional astrophysical effects such as turbulence, three-dimensional asymmetry, and a varying degree of fragmentation cannot be captured by our one zone model. These astrophysical effects are notable features in simulations that use atomic cooling limit as the benchmark threshold for \acro{DCBH} formation, see, for example Refs.~\cite{2019Natur.566...85W, Regan:2020drm}.
We review the status of the field in Section~\ref{sec:simulations:for:collapse}. 

It is also plausible that \emph{in-situ} dark matter decay contributes to \HH{} dissociation, possibly amplified by dark matter adiabatic contraction, as observed in Ref.~\cite{Lu:2024zwa}.
At high densities, however, many ro-vibrational levels in the ground electronic state become populated, so that a narrow photon spectrum may have fewer viable \HH{} dissociation targets.

\paragraph{Is there a critical \HH{} fraction for \acro{DCBH}?}
Our study establish a critical \HH{} fraction, \eqref{eq:.xH2.crit}, for a halo to reach the atomic cooling limit necessary milestone in the formation of a direct collapse black hole. It is not yet established what minimum conditions beyond this milestone are required to ensure heavy seed \acro{DCBH} formation. A plausible second milestone is to reach the `zone of no return' at which point the gas will undergo monolithic collapse~\cite{Omukai:2000ic}. While this target may not be necessary for \acro{DCBH} formation, simulations indicate that this evolution is also characterized by keeping the gas below a critical \HH{} fraction. However, recent high-resolution simulations indicate that these may only populate intermediate mass seeds~\cite{Prole:2023toz}. The dynamics that determine heavy seed formation may lie at a third milestone, tracking the evolution of the gas to very high density, $n_\text{p} \gtrsim 10^9 \, \text{cm}^{-3}$. At these densities, three-body formation channels\footnote{$3\Ha{} \to \HH{} + \Ha{}$ and $\HH{} + 2 \Ha{} \to 2\HH{}$, where the third body is necessary to conserve energy and momentum.} inevitably lead to a rapid increase in the \HH{} abundance. By this stage, however, the gas is optically thick to \HH{} emissions and is heating faster than \HH{} excited states can radiate. Other halo factors are likely to determine whether gas in this regime fragments~\cite{Becerra:2014xea, Prole:2023toz}.
This leads to uncertainty on the impact of \HH{} at the third milestone, which factors are most relevant for fragmentation, and whether fragmentation at this stage will prevent heavy seed formation.

\paragraph{Environmental contributions}

The mass accretion model assumes the  median growth of a dark matter halo observed in simulations, time-averaging over periods of rapid growth, relaxed phases, and major mergers.
One expects that the ensemble of actual high-redshift halos realize significant deviations from this median path; presently, the only way to study these is through simulations. 
For example, a limitation of our approach is that we cannot capture the effects of instantaneous mergers on \acro{DCBH} formation. 
Ref.~\cite{Regan:2019vdf} finds that both the critical Lyman--Werner intensities and the dynamical heating thresholds are significantly relaxed when a halo experiences a major merger just as it approaches the atomic cooling limit.
The $\infty$-galaxy may even be a manifestation of these dynamics~\cite{vanDokkum:2025idp}.

A separate environmental effect is metal pollution. Metals\footnote{In the astronomer's sense: any element that is not hydrogen or helium.} are additional coolants that can cause fragmentation analogous to \HH{}. We have assumed that dark matter halos remain metal free in order to be \acro{DCBH} candidates. The metallicity evolution of early galaxies mainly depends on their accretion and merger history, where a dark matter halo inherits the properties of its progenitors~\cite{Loeb&Furlanetto:2013, OBrennan:2025sft}. Since our scenario suppresses star formation before halos reach the atomic cooling limit, we naturally avoid hereditary metal enrichment. 
However, our semi-analytical analysis does not account for metal pollution from nearby supernova-driven galactic outflows. For example, not all halos are expected to have a history that leads to atomic cooling---even in the presence of axion decay. We leave this aspect of the modeling for future simulations.  

On similar grounds, we also assume that our model halo does not experience X-ray radiation. X-rays have a long mean free path in the \acro{IGM} and boost \HH{} production by injecting free electrons in the system. They are therefore detrimental to the formation of \acro{DCBH} candidates and counteract the effect of Lyman--Werner radiation.

\paragraph{Model limitations} Our semi-analytical model for the dark matter and gas in a halo emphasizes simplicity and interpretability. The one-zone model examines the gas core as effectively isotropic and homogeneous so that its chemo-thermal evolution evolves simply in time. The subsequent halo evolution at high gas density may merit more thorough three-dimensional modeling to address self-shielding.

We further assume a quasi-static core when dynamical heating is the dominant temperature-changing effect~\cite{Visbal:2014nogo, Nebrin:2023yzm, Hegde:2023wxz}. 
We set our critical \HH{} abundance with a conservative halo timescale, $\tauHub$, in \eqref{eq:.xH2.crit} in order to stay within this assumption. The optimistic evolution that uses the free fall time $\tau_\text{ff}$ indicates evolution that encroaches beyond the this assumption, see Fig.~\ref{fig:standard:halo:history:and:thermal.rates} (right).
This ultimately limits the scope of our halo evolution to the atomic cooling threshold rather than the zone of no return~\cite{Omukai:2000ic}.

The function \eqref{eq:MAH} is a single mass accretion history that fits the median halo evolution from an ensemble of simulated halos. The intrinsic scatter of different halos thus contributes a range of halo behaviors that are not captured in our treatment. In reality, halos experience periods of rapid growth, relaxed phases, or other significant deviations from the median path, which need to be modeled through simulations.

\subsection{Implications and Future directions}
The key result of our study is that dark matter decay in the intergalactic medium can produce atomic cooling halos---progenitors of direct collapse black holes---which may, in turn, explain the growing population of observed high-redshift supermassive black holes. 
The simplified modeling used in new physics studies like ours offer clear cause-and-effect interpretations of astrophysical evolution, but cannot capture a range of significant effects that require simulations. 
To resolve the puzzle of supermassive black hole observations, we want to assess the \emph{relative contribution} of dark matter energy injection to the nonlinear astrophysical dynamics captured in simulations. 

This is evident when considering the distribution of halos relative to the median case. If a model predicts \acro{DCBH} candidates in a benchmark halo, then what is the spread among other halos that either fail to form candidates or evolve differently?
Understanding this breadth of behaviors would allow us to assess whether standard astrophysics is sufficient to describe the high redshift supermassive black hole population, or whether additional non-standard energy injection is required. 
Although this presents both a computational and an observational challenge, it is a necessary step towards determining whether \acro{JWST} observations can be interpreted as evidence for new physics.

A complementary approach is to ask whether the dynamics described in this paper could be used to set an upper bound on axion--photon couplings for the masses considered. For example, one could imagine an extreme scenario where the Lyman--Werner flux is so potent that \HH{} is suppressed everywhere and \emph{only} the first galaxies \emph{all} begin as atomic cooling halos or even direct collapse black holes. This scenario appears unlikely without more sophisticated modeling, both because of the range of astrophysical dynamics and because of the proximity of existing axion bounds to the coupling strengths needed to induce \acro{DCBH} candidates.

With this broad vision in mind, we identify promising directions that merit further inquiry.

\paragraph{Subsequent evolution, population studies, simulations}
Presently, all astrophysical simulations of heavy seed formation imagine that Lyman--Werner radiation comes from nearby star-forming galaxies. This paper and its predecessors---Refs.~\cite{Friedlander:2022ovf,Lu:2024zwa} by others---motivate the role of including a dark matter contribution in these studies. The nature of this flux offers different spectra, morphology, and time evolution compared to the standard assumptions.  These studies would be an impactful bridge between dark matter studies and simulations that could then permit more quantifiable studies of the \acro{DCBH} population in the presence of dark matter energy injection.

\paragraph{Dark matter models} 
Our axion model is readily adapted to any dark matter model decaying into a pair of photons. Earlier work examined non-monochromatic spectra such as three-photon decays using a parabolic spectrum~\cite{Friedlander:2022ovf,Lu:2024zwa}. These studies treated \emph{in situ} decays using a cross section across the Lyman--Werner band, see Appendix~\ref{sec:constant:LW:cross:section}, and may merit a re-investigation given the impact of the Lyman--Werner line structure in this paper.

One may also combine the effects of halo adiabatic contraction in Ref.~\cite{Lu:2024zwa} with the intergalactic medium contribution in this paper---potentially suppressing \HH{} formation during the entire period from pre-virialization to post-atomic cooling. 
This may require some care to account for the Lyman--Werner line structure for the \emph{in situ} contribution. It may be, for example, that the \emph{in situ} contribution is only effective at large densities when \HH{} populates rotational excitations of the \HH{} ground state levels---in that case, the number of Lyman--Werner lines dramatically increases and may be reliably treated as a band.

One may also consider dark matter annihilation into \acro{UV} photons. It is more challenging to develop models of annihilating dark matter in this mass range that realize the observed dark matter abundance. This challenge notwithstanding, one may examine the relative contribution of a dark matter annihilation from the intergalactic medium compared to within the halo. Unlike decaying dark matter, the flux depends on the square of the dark matter density, which favors the halo contribution.

Finally, one may engineer photon spectra with more complicated dark sectors. One possible reason to do this would be to provide a background that rapidly  excites the \HH{} rotational ground states. In this case, the continuum Lyman--Werner cross section approximation is valid and one may potentially expand the model reach along the lines of Fig.~\ref{fig:counterfactual} (left). A second reason to engineer spectra is in concert with very different dark matter models. For example, the dark star scenario proposes that $\mathcal O(100\,\text{GeV})$ dark matter annihilation could support stars by providing pressure support against standard cooling mechanisms~\cite{Spolyar:2007qv,Natarajan:2008db,Spolyar:2009nt}. One could imagine that a \acro{UV} flux that suppresses \HH{} enough to lower the barrier for dark star formation at cosmic dawn---these could even contribute to heavy seed formation~\cite{Banik:2016qww}.

\paragraph{Delayed star formation} 
In our scenario, when the photon flux from axion decay is unable to suppress \HH{}, the halo is assumed to produce \popiii{} stars from \HH{} cooling. These stars, however, are delayed relative to standard cosmology by an amount dependent on the Lyman--Werner flux, see~Fig.~\ref{fig:histories}. 
This delayed \popiii{} formation may itself be a distinguishable signature of the stellar initial mass function as \acro{JWST} seeks to observe the first stars.

\paragraph{Cosmic dawn} The hydrogen 21\,cm signal is one of the most promising windows into cosmic dawn. 
The key quantity is the spin temperature which sets the relative abundance of hyperfine triplet states to the singlet state.
The spin temperature couples strongly to Lyman-$\alpha$ photons via the Wouthuysen--Field effect, and to the \acro{IGM} temperature through heating~\cite{Loeb&Furlanetto:2013}. 
Axion decay may modify the spin temperature:
\begin{enumerate}
    \item Axions between 
    $10.2 \,\text{eV} 
    \leq  \tfrac{m_a}{2} < 
    12.09 \,\text{eV}$ 
    produce photons that redshift into the Lyman-$\alpha$ band. 
    Axions in the narrow band
    $12.09 \, \text{eV} 
    \leq \tfrac{m_a}{2} < 
    12.75 \, \text{eV}$ 
    do not produce Lyman$-\alpha$ photons from their allowed transitions, $3\text{p} \to 1\text{s} \text{ or } 3\text{p} \to 2\text{s}$.

    \item Direct collapse black holes from our scenario may produce X-rays that heat, ionize, and produce its own source of Lyman-$\alpha$ photons. 

    \item Delayed star formation due to an \HH{} suppressing background also delays early sources of \acro{UV} radiation that is released into the intergalactic medium.  
\end{enumerate}
These effects plausibly contribute to the 21\,cm signal and merit further investigation.

\section*{Acknowledgments}

We thank
Anson D'Aloisio, 
Melissa Diamond, 
Avi Friedlander, 
Steve Furlanetto,
Cosmin Ilie,
Yifan Lu, 
Katie Mack, 
Julian Muñoz, 
Priya Natarajan, 
Zack Picker,
Wenzer Qin, 
Fred Rasio, 
Pearl Sandick, 
Sarah Schön, 
Tracy Slatyer, 
and
Aaron Vincent
for many insightful comments and discussions.\footnote{\acro{PT} especially acknowledges Sarah Schön for foundational discussions that motivated this project, and Priya Natarajan for encouragement and providing libations  to the participants of the \acro{ACP} ``Progress after Impasse: New Frontiers in Dark Matter'' workshop.} 
We acknowledge Hai-Bo Yu for his regular encouragement.
This work was performed in part at 
the Aspen Center for Physics (grant \acro{NSF PHY-2210452}), 
the Kavli Institute for Theoretical Physics (\acro{NSF} grant \acro{NSF PHY-2309135}), 
the Mitchell Institute at Texas A\&M University, 
the Center for Theoretical Underground Physics and Related Areas (\acro{CETUP*}) / The Institute for Underground Science at Sanford Underground Research Facility (\acro{SURF}, supported by the South Dakota Science and Technology Authority),
the 2024 \emph{Dark Matter, First Light} conference at the Perimeter Institute for Theoretical Physics,
and the Fermilab Theory Division through the Summer Visitor Program. 
The authors thank these precious institutions for the crucial role they play in enabling theoretical research.
\acro{PT} is supported by a \acro{NSF CAREER} award (\#2045333). \acro{JBD} is supported by the National Science Foundation
under grant~\acro{PHY-2412995}. \acro{TX} is supported by the National Science Foundation
under grant~\acro{PHY-2412671}.

\appendix
\part*{Appendices}

\section{Halo Model}
\label{sec:AP:standard:halo:dynamics}

We describe the time evolution and internal dynamics of the model halo that we use in our analysis. This appendix collects results from prior studies to keep the present paper self-contained. In this appendix, $h = 0.67$ refers to the dimensionless Hubble constant normalized to $H_0= $ 100\,km/s/Mpc~\cite{Planck:2018vyg}.

\subsection{Mass Accretion History}
\label{sec:halo:time:evolution}

In the \acro{$\Lambda$CDM} cosmological paradigm, structure forms hierarchically: smaller structures collapse first and larger structures are formed later through accretion and mergers.
The size of the collapsed structure is set by a balance between the effects of gravity and pressure. 
Being pressureless, dark matter quickly collapses into gravitationally bound, planet-sized proto-halos. Baryons, on the other hand, do not collapse until the halos are massive enough, $\mathcal O(10^5\,\text{M}_\odot)$, to overcome the gas pressure.

The mass accretion history formalism describes the total mass of the halo as a function of time, $M(z)$. In \eqref{eq:MAH} we invoke the fitting function presented in Ref.~\cite[App.~B]{Correa:2014xma}.\footnote{Ref.~\cite{Correa:2014xma} implicitly writes all masses as dimensionless numbers in units of solar mass, $M_\odot$.} 
The model parameters $p$ and $f(M_0)$ depend on the growth factor and variance of the cosmic density field, respectively. 
We generate the cosmic density variance with the Python package \texttt{HaloMod}~\cite{Murray:2013qza,Murray:2020dcd} matched to Planck cosmology~\cite{Planck:2018vyg}.

\subsection{Gas virialization}
In the mass range for this paper, axion decays neither ionize nor heat the gas. The gas dynamics therefore proceed independently of the axion decays. 
A cloud of gas collapses at the Jeans threshold, when its pressure can no longer support it against gravity. A related scale, the filtering mass, accounts for the relative velocity of the baryons and dark matter~\cite{Gnedin:1997td,Gnedin:2025opp}. 
In this approach, the halo virializes when the total mass matches the filtering mass~\cite[eq.~8]{Hegde:2023wxz}, $M(z_\text{vir}) = M_\text{F}(v,z_\text{vir})$, where
\begin{align}
    M_\text{F}(v,z) 
    &\approx
    \left(1.66 \times 10^4  \, M_\odot \right)
    \left( 
        1 
        + 
        \frac{ v }{ \sigma_v(z) } 
        \right)^{\!5.02}
    \left( 
        \frac{ 1 + z }{ 21 }
    \right)^{\!0.85} \ , 
\end{align}
were $v$ is the relative (stream) velocity between dark matter and baryons. We assume $v$ follows a Maxwellian distribution with a root-mean-square value of $\sigma_v(z) = 3 \times 10^{-2} (1+z) \, \text{km} \, \text{s}^{-1}$   ~\cite{Tseliakhovich:2010bj, Nebrin:2023yzm, Hegde:2023wxz}. In \eqref{eq:halo:model} we assume $v/\sigma \to 0$, which conservatively allows \HH{} formation to begin as soon as possible.

\subsection{Gas morphology and temperature}

Prior to virialization, $z>z_\text{vir}$, we assume that the gas follows the intergalactic medium density and temperature. Simulations show that after virialization and in the absence of cooling mechanisms, the gas within the halo is suspended in a quasi-static state: it forms a central core with constant density, surrounded by an outer enve{}lope where the gas density falls off with radius as $r^{-2}$~\cite{Machacek:2000us, Wise:2007jc, Visbal:2014nogo, Inayoshi:2015yqa, Chon:2016nmh, Nebrin:2023yzm, Hegde:2023wxz}. The core size is approximately a tenth of the virial radius~\cite{Visbal:2014nogo}, 
\begin{align}
 R_\text{core}(z) &= 0.1\,R_\text{vir}(z)   
 \label{eq:Rcore}
 \ .
\end{align}
We model the gas core density $n_\text{p}$ and temperature $T$ by piecewise-matching the pre-collapse intergalactic medium values to the quasi-static values, \eqref{eq:halo:gas:temp:patch}. These are shown in Fig.~\ref{fig:background:halo:evolution}.

\paragraph{Morphology}

The post-virialization core gas density depends on the halo mass. Simulations show that halos with masses less than $10^6 \ M_\odot$ have a core density that scales like a power of the virial temperature, $T_\text{vir}^{3/2}$ while larer halos have a constant core density~\cite{Visbal:2014nogo}. We match our virialized core density to these simulations: 
\begin{align}
    n_\text{core}(z) = 
    \text{Min}\!\left[\; 
        \frac{5.55}{\text{cm}^3}
        \left(
            \frac{ T_\text{vir} }{ 1000\,\text{K} }
        \right)^{3/2}
        \; , \;\;
        \frac{3.24}{\text{cm}^3}
    \;\right] \ .
    \label{eq:n:core}
\end{align}
While we do not use it in our one-zone analysis, we note that the density profile away from the core scales as $[1+(r/R_\text{core})^2]\inv$~\cite[eq.~5]{Nebrin:2023yzm}.
The pre-virialization density follows from redshifting the present intergalactic medium proton density,
\begin{align}
    n_{\text{p}0} 
    = 
    1.9 \times 10^{-7} \, \text{cm}^3 
    \ .
\end{align}

\paragraph{Temperature} The gas (matter) temperature in the intergalactic medium is modeled with \texttt{CosmoRec}~\cite{Chluba:2010ca,Chluba:2010fy}, see also Ref.~\cite[Fig.~2.5]{Loeb&Furlanetto:2013}. 
\begin{align}
    T_\text{IGM} (z) = 0.025 
    (1+z)^2 
    \, \text{K} 
    \ . 
\end{align}
After virialization, the gas undergoes shock heating and adiabatic contraction, reaching the virial temperature~\cite[eq.~3.32] {Loeb&Furlanetto:2013}, see also \cite{Tegmark:1996yt, Bromm:2001bi},
\begin{align}
    T_\text{vir} (M,z) 
    & = 
    1.48 \times 10^4 \, \text{K} \;
    \left( \frac{\mu}{0.6}\right)
    \left(
    \frac{\Omega_\text{m0}}{\Omega_\text{m}(z)}
    \frac{\Delta_c}{18 \pi^2}
    \right)^{1/3}
    \left(\frac{M}{10^8 h^{-1} M_\odot}\right)^{2/3}
    \left( \frac{1+z}{10}\right) \ ,
    \label{eq:T.vir}
\end{align}
where $\mu = 1.22$ is the mean molecular mass of a the primordial gas mixture in units of the \Ha{} mass, $\Delta_\text{c}$ is the concentration parameter \eqref{eq:concentration:parameter}, and $\Omega_\text{m}(z)$ is the matter density as a fraction of the critical density with $\Omega_{\text{m}0}=\Omega_\text{m}(0)$. An additional factor of $0.75$ is included to better match the simulation results, as suggested in Ref.~\cite{Fernandez:2014wia}.
In the absence of a cooling mechanism, the halo's temperature continues to follow \eqref{eq:T.vir} as it grows via accretion and mergers~\cite{Yoshida:2003rw, 2019Natur.566...85W}. 

\paragraph{Model validity} A key assumption in this model is the use of the quasi-static core in the limit where dynamical heating is the only relevant temperature-changing process~\cite{Visbal:2014nogo, Nebrin:2023yzm, Hegde:2023wxz}. 
At the onset of a cooling mechanism---when $\tau_\text{cool} \sim \tau_\text{dyn.}$---the model is no longer appropriate. We remark that this transition may be a topic that merits more detailed simulation.

\subsection{Parameterizing Self Shielding}
\label{app:self:shielding}

Self-shielding is the phenomenon the build up of molecular hydrogen in a halo further protects the core of the halo from \HH{}-dissociating Lyman--Werner radiation. Thus there can be a snowball effect where a small amount of \HH{} leads to a positive feedback loop that facilitates further \HH{} production. This phenomenon is included as a self-shielding factor, $f_\text{sh}$, in the photo-dissociation rate $k_\text{LW}$ in \eqref{eq:LyWer:k:numeric}. 

\paragraph{Fitting function}
Draine and Bertoldi introduced a fitting function for $f_\text{sh}$~\cite{1996ApJ...468..269D}
\begin{align}
    f_\text{sh}( y , T ) 
    &= 
    \frac{
        a_1
        }{ 
        \left( 
            1 + b_5\inv y
        \right)^{\eta}
        }
    + 
    \frac{
        a_2
        }{
        \left( 
            1 + y
        \right)^{0.5} 
        }
    \,
    \exp{\!
        \left[
            - 
            a_3
            \left( 1 + y \right)^{0.5}
        \,\right] 
        } 
    \ ,
    \label{eq:f.sh}
\end{align}
where we use a dimensionless measure of the \HH{} column density $N^\text{col}_\HH{}$,
\begin{align}
    y = 
    \frac{ N^\text{col}_\HH{} }{ 5 \times 10^{14}\, \text{cm}^{-2} }
\end{align}
and we list the dimensionless $a_i$ and $b_5$ coefficients in the first row of Table~\ref{table:shielding:parameters}. $b_5 = b/(10^5\,\text{cm}\,\text{s}\inv)$ is a dimensionless measure of the full-width-at-half-maximum Doppler broadening parameter, $b=\text{\acro{FWHM}}/\sqrt{4}\ln 2$.  The original fit found $\eta = 2$, with a later revision in Ref.~\cite{Wolcott-Green:2011tul} finding a better agreement (within a factor of 2) with simulations with $\eta= 1.1$ for the temperature range $500\,\text{K} < T < 5000\,\text{K}$ and column densities 
$y \leq 10^{6}$.
Wolcott--Green and Haiman later included the rotovibrational distribution of \HH{}, which is itself a function of the gas density $n$ and temperature $T$~\cite{Wolcott-Green:2018hyx}. They find that promoting $\eta$ to a function improves the fit to the percent level $n \leq 10^7\, \text{cm}^{-3}$, $T\leq 8000\,\text{K}$, and $y \leq 10^{3}$:
\begin{align}
    \eta(n, T) 
    &= 
    A_1(T) 
    \exp{\left( 
        - c_1 \log_{10}{ \frac{n}{ \text{cm}^3 } } 
        \right) } 
    + 
    A_2(T) \label{eq:eta} 
\end{align} 
The temperature-dependent functions are
\begin{align}
    A_1(T) 
    &= c_2 \log_{10}\left(\frac{T}{\text{K}}\right) - c_3 
    & 
    A_2(T) 
    &= - c_4 \log_{10}\left(\frac{T}{\text{K}}\right) + c_5
    \ .
\end{align}
We list the dimensionless $c_i$ coefficients in the second row of Table~\ref{table:shielding:parameters}.

\begin{table}
    \renewcommand{\arraystretch}{1.3} 
    \centering
    \begin{tabular}{ p{2cm} | p{2cm}p{2cm}p{2cm}p{2cm}p{2cm}  } \toprule
        Ref.~\cite{1996ApJ...468..269D}
        & $a_1$ & $a_2$ & $a_3$ & $b_5$&
        \\ \hline
        & $0.965$ & $0.035$ & $8.5\times 10^{-4}$& $3.0$ &
        \\ \midrule
        Ref.~\cite{Wolcott-Green:2018hyx} 
        & $c_1$ & $c_2$ & $c_3$ & $c_4$ & $c_5$
        \\ \hline
        & $0.2856$ 
        & $0.8711$
        & $1.928$
        & $0.9639$ 
        & $3.892$ 
        \\ \bottomrule
    \end{tabular}
    \caption{
        Dimensionless fitting parameters for the shielding factor $f_\text{sh}$. 
        \label{table:shielding:parameters}
  }
\end{table}

\paragraph{Parameters for protogalaxies}
The column density $N^\text{col}_\HH{}$ is the \HH{} number density multiplied by an appropriate length scale; see, e.g., Ref.~\cite{Wolcott-Green:2011tul}. 
Approximate this with the \HH{} number density at the core~\cite{Nebrin:2023yzm}, $x_\HH{} \, n_\text{core}$ with the core radius 
\eqref{eq:Rcore}
and core density \eqref{eq:n:core},
\begin{align}
    y = 0.926 \, 
        \frac{ 
            n_\text{core}\,x_\HH{}\, R_\text{core}
        }{ 
            5 \times 10^{14}\, \text{cm}^{-2} 
        } \ .
\end{align}

\section{Comparison to Local Contribution}
\label{sec:local:contribution}

We review the argument by Qin~et\ al.\ that the energy injected from dark matter decays in the intergalactic medium can be much larger than those from within the halo (\emph{in situ})~\cite{Qin:2023kkk}. In so doing, we present formulae for dark matter distribution in the halo. While our primary study does not require  information this distribution directly, it relates to our working definition of the halo virial radius that, in turn, defines our characteristic core size. 

\subsection{Dark Matter Distibution}
\label{eq:internal:halo:structure}

We assume that axion dark matter follows an \acro{NFW} profile~\cite{1997ApJ...490..493N}
\begin{align}
    \rho_a(r, z) 
    &= 
    \frac{ \rho_0 (z) }{ 
        \frac{ c_\text{h} r }{ R_\text{vir} (z) }
        \left( 1 + 
            \frac{ c_\text{h} r }{ R_\text{vir}(z) }
        \right)^2
        } 
    &
    \rho_0 (z) 
    &= 
    \left(
        \frac{ M(z) }{ 4 \pi R_\text{vir} (z)^3 }
    \right)
    \frac{ c_\text{h}^3 }{ 
        \ln{ (1 + c_\text{h} ) } 
        - 
        \frac{c_\text{h}}{ 1 + c_\text{h}  } 
    }
        \ ,
\end{align} 
The choice of the concentration parameter $c_\text{h}=4$ is appropriate for small proto-halos undergoing rapid growth \cite{Zhao:2008wd}. 
The virial radius is, see e.g.\ Ref.~\cite[eq.~3.30]{Loeb&Furlanetto:2013}, 
\begin{align}
    R_\text{vir}(M, z)
    &= 
    (784~\text{pc})
    \left(
        \frac{ \Omega_{\text{m}0} }{ \Omega_\text{m}(z) } 
        \frac{ \DeltaC{} }{ 18 \pi^2 }
        \right)^{-1/3}
    \left(
        \frac{ M(z) }{ 10^8\, h^2 \, M_\odot  } 
    \right)^{1/3}
    \left( 
        \frac{1+z}{10}
    \right)^{-1} \
    \ .
    \label{eq:virial.radius}
\end{align}
The matter density parameter and mean overdensity at collapse are
\begin{align}
  \Omega_\text{m}(z) 
  &= 
  \frac{ 
    \Omega_{\text{m}0} (1+z)^3 
    }{ 
    \Omega_{\Lambda 0} + \Omega_{\text{m} 0} (1+z)^3 
   } 
   &
    \DeltaC{}
   &= 
   18\pi^2 
   + 82d 
   - 39d^2 
   \approx 200
   \label{eq:concentration:parameter}
   \ ,
\end{align}
where $d= \Omega_\text{m}(z_\text{col}) - 1$ is evaluated at the redshift of collapse, $z_\text{col}$.
$\DeltaC{}$ is the ratio of the virial mass of the halo to a ball of radius $R_\text{vir}$ with density $\rho_\text{crit}$.

\begin{figure}[tb]
    \centering
    \includegraphics[width=\linewidth]{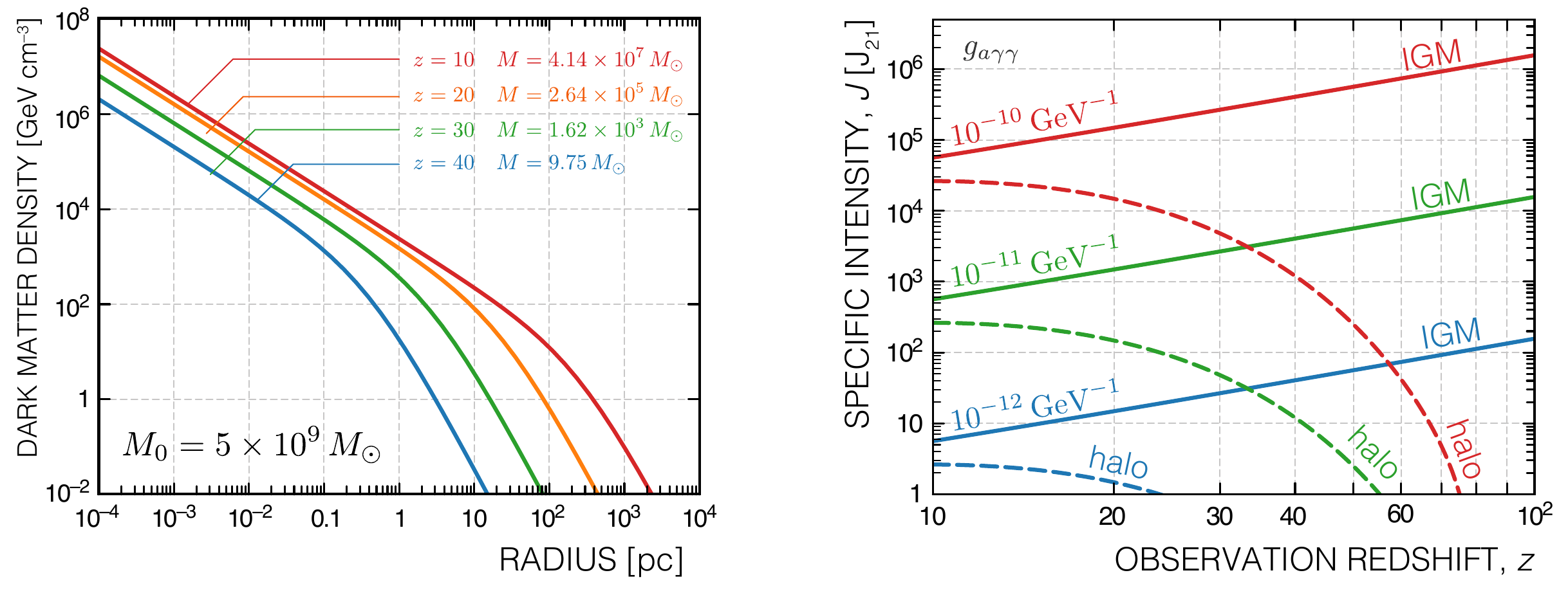}
    \caption{%
    \textsc{Left}: Internal structure of our model dark matter halo as a function of time, using the mass accretion history prescription~\cite{Correa:2014xma}. The halo parameter is $M_0 = 5\times 10^ 9\,M_\odot$, and it reaches the atomic cooling limit at $z=10$.
    \quad 
    \textsc{Right}: 
    Comparison of \emph{in situ} (solid) and intergalactic medium contributions to the halo flux without including any flux attenuation from Lyman lines. For the \emph{in situ} contribution we assume an intensity that is equally spread across the Lyman--Werner band.}
    \label{fig:Jigm:halo}
\end{figure}

\subsection{Local versus external flux}

The halo contribution to the flux is
\begin{align}
    J_\text{halo} 
    & = 
    \frac{h}{4 \pi} 
    \frac{ \Gamma_a }{ m_a } 
    E_\gamma
        \frac{ \D{N} }{ \D{E_\gamma} }
    D_\text{halo} 
    &
    D_\text{halo} 
    &= 
    \int_{R_\text{min}}^{R_\text{vir}} \D{s}\; 
    \rho_{a}(s)
    \ .
    \label{eq:J.halo}
\end{align}
The $D$-factor encompasses the geometry of the halo and $\D{N_\gamma} / \D{E_\gamma}$ is the photon spectrum from axion decay. 
The integral over the \acro{NFW} profile diverges logarithmically as the $R_\text{min}\to 0$. The $D$-factor is insensitive to a short distance cutoff so we may set $R_\text{min} = 10^{-4} \, \text{pc}$. 
Fig.~\ref{fig:Jigm:halo} plots the \emph{in situ}  photon flux compared to that of the intergalactic medium as a function of redshift. The latter dominates until at $z=10$, which is the endpoint of our model halo evolution.

\subsection{Doppler broadening}
Ref.~\cite{Lu:2024zwa} suggests that Doppler broadening may spread out the photon line spectrum from axion decay. This is critical for the \emph{in situ} contribution to overlap with one of the Lyman--Werner bands of the \HH{} molecule.
An axion temperature can be defined using the virial theorem~\cite[eqs.\,3.30-32]{Loeb&Furlanetto:2013},
\begin{align}
    T_{a} 
    &= 
    \frac{ m_{a} V_{c}^2 }{2 k_\text{B}} 
    &
    V_\text{c} 
    &= 
    \left(
    \frac{G M}{R_\text{vir}}
    \right)^{1/2}
    &
    \sigma_\text{V} 
    &= 
    \left( 
        \frac{k_\text{B} T_a }{ m_a }
    \right)^{1/2} 
    = \left( 
        \frac{G M}{2 R_\text{vir}}
        \right)^{1/2} 
    \label{eq:dispersion.velocity}
    \ .
\end{align}
$V_\text{c}$ is the axion circular velocity and $\sigma_\text{V}$ is the velocity dispersion~\cite[\S6.5]{Draine:2011ism}. 
The full width at half maximum (\acro{FWHM}) is
\begin{align}
    (\Delta E)_\text{FWHM} = 
    m_a \sqrt{\ln{2}}
    \frac{ V_\text{c} }{c} \ .
\end{align}
In early halos, the velocity dispersion is on the order of up to $\mathcal O(10 \, \text{km}\,\text{s}^{-1})$ \cite[Fig.~3.6]{Loeb&Furlanetto:2013}. This corresponds to energy shifts of $ \Delta E_\gamma \lesssim  10^{-4} m_a$.
An axion with mass $m_a = 25 \, \text{eV}$ in a $10^8\,M_\odot$ halo will have $V_\text{c} \sim 30 \, \text{km/s}$ and a spectral broadening of $(\Delta E)_\text{FWHM} = 2.1\times 10^{-3} \, \text{eV}$. This much smaller than the gap between Lyman--Werner bands---see Fig.~\ref{fig:Lyman.Werner.bands}---and is effectively a line. This means that the \emph{in situ} contribution to photodissiciation at low density requires the axion mass to be tuned to a Lyman--Werner transition.

\section{Molecular Hydrogen and Photodissociation}
\label{AP:H2:molecule}

We review the basic molecular chemistry of \HH{} molecule and present our treatment of the Lyman--Werner lines compared to the standard constant cross section approximation. 

\HH{} is a diatomic molecule with a binding energy of $\sim 4.48 \, \text{eV}$. Because it is homonuclear, it has no electric dipole moment and is thus stable against $\HH{} + \gamma \to 2\Ha{}$. This means that the destruction of \HH{} must first proceed by lifting the ground state to an electronic excited state. The excited state may then subsequently decay to the antisymmetric (anti-bonding) unbound state. 

\subsection{Term symbol and rotovibrational notation}

This section draws from \emph{Physics of the Interstellar and Intergalactic Medium} by Draine~\cite[\S5.1.3]{Draine:2011ism}.
Molecular energy levels are designated by \emph{term symbols} $\mathcal{Z}^{(2 S + 1)} \mathcal{L}_{u,g}$ where $S$ is the total spin, $\mathcal L = \Sigma, \Pi$ respectively represent orbital angular momenta projections of 0 and 1 along the internuclear axis, and $\mathcal{Z} = \text{X}, \text{A}, \text{B}, \text{C}$ distinguish the ground (X) from excited states. 
The subscripts $g$ (gerade) or $u$ (ungerade) specify whether the wave function is parity even or odd.
The $\mathcal{L} = \Sigma$ has a degeneracy is labeled by a $\pm$ superscript indicating symmetry or antisymmetry through a plane containing the internuclear axis. The first two sets of electronic excitations of the \HH{} molecule from its ground state are
\begin{align}
    &
    \text{X}^1\Sigma^+_g 
    \to
    \text{B}^1 \Sigma^+_u \ 
    (\text{Lyman})
    &
    \text{X}^1\Sigma^+_g 
    \to
    \text{C}^1 \Pi_u \ 
    (\text{Werner})
    \label{eq:Ly:Wer:transitions}
\end{align}
These are also written $\text{B}(v, J)$ and $\text{C} (v,J)$ to indicate the rotovibrational quantum numbers. 
For convenience, we label the ground states by $\alpha \in \text{X}(v,J)$, and excited states by $\dot{\alpha} \in \text{B}(v',J') \ \text{or} \ \text{C} (v',J')$.
In low density environments, $\mathcal O(\text{cm}^{-3})$, \HH{} typically exists the $v=0$ and $J = 0,1$ (ortho-\HH{} and para-\HH{}) electronic ground state with a relative abundance of 3:1. The excitations \eqref{eq:Ly:Wer:transitions} represent $\mathcal O(70)$ transitions in the $11.18 - 13.6 \ \text{eV}$ energy range~\cite{1989A&AS...79..313A,Haiman:1999mn, Ahn:2008uwe, 2019ApJ...885..163X} (see the Appendix of \cite{Haiman:1999mn} for exact structure). 

\subsection{Lyman--Werner Band Structure}
\label{AP:Lyman:Werner:Bands}

Electric dipole selection rules permit transitions with $J\to J \pm 1$, called the $R$ and $P$ branches, or $J\to J$, called the $Q$ branch.  Specific transitions are labeled $R(J)$, $P(J)$, and $Q(J)$. The Werner $P(1)$ transition and any non-Werner $Q(J\neq 0)$ transitions are forbidden.
We use a notation were, for example, Ly-$R$(0) corresponds to a transition $\text{X}(0,0) \to \text{B}(v',1)$, and Wr-$Q$(1) denotes $\text{X}(0,1) \to \text{C}(v',1)$, for an vibrational state $v'$.

\begin{figure}[tb]
    \centering
    \includegraphics[width=\linewidth]{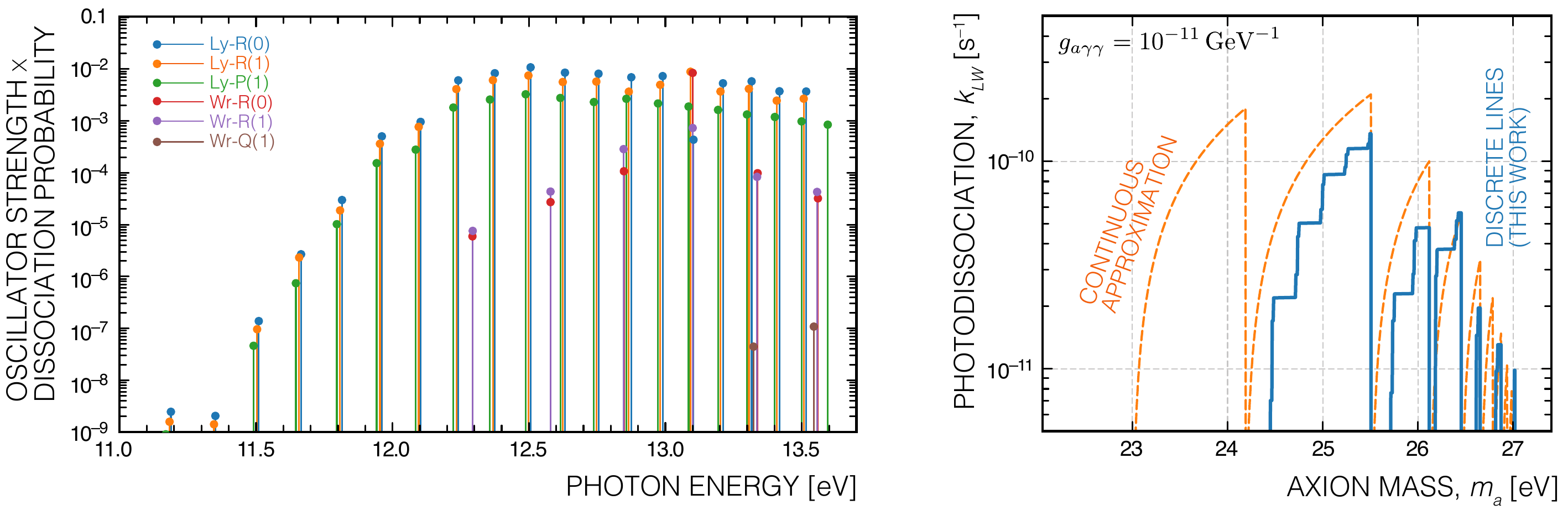}
    \caption{%
    \acro{Left}:
    All available Lyman and Werner transitions for the \HH{} molecule below the Lyman limit, for ortho-\HH{}, X$(0,1)$, and para-\HH{}, X$(0,0)$. 
    \quad 
    \acro{Right}:
    Lyman--Werner rate coefficient as a function of axion mass for the continuum Lyman--Werner band approximation (dotted) and our treatment of the Lyman--Werner line structure (solid). The sawtooth structure represents Lyman lines in atomic hydrogen, while the difference from the dotted to solid line represent the gaps between the Lyman--Werner transitions of molecular hydrogen.
    }
    \label{fig:Lyman.Werner.bands}
\end{figure}
We compile the data for the Lyman system~\cite{1993A&AS..101..273A}, the Werner system~\cite{1993A&AS..101..323A}, and the dissociation probabilities~\cite{1992A&A...253..525A}, to calculate an effective dissociation oscillator strength analogous Ref.~\cite[App.]{Haiman:1999mn}. 
The oscillator strength times the dissociation probability is \cite[eq.~6.20]{Draine:2011ism} (see also \cite{1989A&AS...79..313A}),
\begin{align}
    f_{\alpha \dot{\alpha}} 
    \text{P}_{\dot{\alpha}}^\text{dis}
    & = 
    2.31 \times 10^{-8} 
    \frac{(2 J_{\dot{\alpha}} + 1)}{(2 J_\alpha + 1)}
    \left(\frac{A_{\dot{\alpha}\alpha}}{\text{s}^{-1}} \right)
    \left( \frac{E_{\dot{\alpha}\alpha}}{\text{eV}}\right)^{\!-2}
    \text{P}_{\dot{\alpha}}^\text{dis}
    \label{eq:foscillator:Pdis}
    \ ,
\end{align}
where $J_\alpha$ and $J_{\dot{\alpha}}$ are the rotational quantum numbers in the ground and excited state respectively, $A_{\dot{\alpha} \alpha}$ is the Einstein $A$ coefficient for the $\dot{\alpha} \to \alpha$ transition, and $E_{\dot{\alpha} \alpha}$ is the transition energy between them. We plot \eqref{eq:foscillator:Pdis} in Fig.~\ref{fig:Lyman.Werner.bands}, shows the available Lyman and Werner dissociation lines that are available below the Lyman limit.

\subsection{Constant Cross Section Approximation}
\label{sec:constant:LW:cross:section}

When the gas density and temperature increases, the \HH{} ground state populates its higher ro-vibrational modes beyond the ortho and para approximation, $v \neq 0$ and $J \neq 0,1$. The number of transitions \eqref{eq:Ly:Wer:transitions} grows to $\mathcal O(10^5)$~\cite{Shaw:2005xm, Wolcott-Green:2018hyx} and one may be better served with a continuum approximation.  

We integrate \eqref{eq:LyWer:k:full:expression} with a constant cross section~\cite[Sec.~6.1.1]{Loeb&Furlanetto:2013}
\begin{align}
    \Bar{\sigma}_\text{LW} 
    &= 
    3.71 \times 10^{-18} \, \text{cm}^2 
    &
    & 
    11.5 \, \text{eV} \;\leq h\nu <\; 13.6 \, \text{eV}\label{eq:sigma:LW:constant}
    \ ,
\end{align}
which is derived by averaging over the $\sim \mathcal{O}(70)$ lines in the \HH{} ground state. We plot the resulting photodissociation flux compared our treatment of the line spectrum in Fig.~\ref{fig:Lyman.Werner.bands}. The impact of this approximation is seen by comparing the axion parameter plot with the continuous cross section approximation Fig.~\ref{fig:counterfactual} (left) with our main result, Fig.~\ref{fig:Axion.DCBH}.

The constant cross section approximation is common for  astrophysical sources of Lyman--Werner radiation that typically have broad spectra. As we highlight in this study, the approximation is not appropriate for spectra that are narrow compared to the Lyman--Werner band.
However, large \HH{} number densities and temperatures cause the ground-state \HH{} molecules to populate their $v\neq 0$ rotational modes which are then sensitive to  $\mathcal O(10^5)$ transitions. This is much larger than the $\mathcal O(70)$ transitions from the rotational ground state in Fig.~\ref{fig:Lyman.Werner.bands}. In this regime, it may be more appropriate to take the continuum approximation~\cite{Wolcott-Green:2011tul,Wolcott-Green:2016grm,Wolcott-Green:2018hyx}, though one should be careful to average over the larger population of lines rather than just the rotational ground state lines. The regime where the constant cross section approximation is valid may coincide with the regime where adiabatic contraction boosts the \emph{in situ} axion decay contribution~\cite{Lu:2024zwa}.

\bibliographystyle{utcaps} 	
\bibliography{DCBH}

\end{document}